\documentclass[11pt]{article}

\pdfoutput=1

\usepackage{graphicx, color, url}
\usepackage{authblk}
\usepackage[colorlinks,citecolor=blue,urlcolor=blue,linkcolor = blue]{hyperref}
\usepackage{amsmath}
\usepackage{dsfont}
\usepackage{amssymb}
\usepackage{epstopdf}
\usepackage{boxedminipage}
\usepackage{amsfonts}
\usepackage{url}
\usepackage{amsbsy}
\usepackage{setspace}
\usepackage{natbib}
\usepackage{bm}
\usepackage{wasysym}
\usepackage{multirow}
\usepackage{textcomp}
\usepackage{amsthm}
\usepackage{verbatim}
\usepackage{enumerate}
\usepackage{amssymb}
\usepackage{amsmath}
\usepackage{amsbsy}
\usepackage{setspace}
\usepackage{natbib}
\usepackage{bm}
\usepackage{wasysym}
\usepackage{textcomp}
\usepackage{amsmath,amssymb,amsfonts, amsbsy, epsfig,natbib,soul, mathtools}
\usepackage{subfig}
\usepackage{mathrsfs}
\usepackage{algpseudocode}
\usepackage{algorithm}
\usepackage{graphicx, color, url}
\usepackage{amsmath}
\usepackage{enumitem}
\usepackage[titletoc,toc,title]{appendix}
\newtheorem{remark}{Remark}
\newtheorem{theorem}{Theorem}

\newtheorem{proposition}{Proposition}
\newtheorem{lemma}{Lemma}
\newtheorem{definition}{Definition}

\title{A Unified Multiple Testing Framework based on $\rho$-values}
\author[$\dag$]{Shenghao Qin}
\author[$\dag$]{Bowen Gang}
\author[$\dag$]{Yin Xia\thanks{Corresponding author: xiayin@fudan.edu.cn}$^{,}$}
\affil[$\dag$]{\it Department of Statistics and Data Science, Fudan University}
\date{}

\begin{document}
	\maketitle

\begin{abstract}
Multiple testing is an important research area with widespread scientific applications, including in biology and neuroscience. Among popularly adopted multiple testing procedures, many are based on p-values or Local false discovery rate (Lfdr) statistics. However, p-values—often obtained via the probability integral transform of standard test statistics—typically lack information from the alternatives, resulting in suboptimal performance. In contrast, Lfdr-based methods can achieve asymptotic optimality, but their ability to control the false discovery rate (FDR) hinges on accurate estimation of the Lfdr, which can be challenging, especially when incorporating side information. In this article, we introduce a novel and flexible class of statistics, termed $\rho$-values, and develop a corresponding multiple testing framework that integrates the strengths of both p-values and Lfdr, while addressing their respective limitations.  Specifically, the $\rho$-value framework unifies these two paradigms through a two-step process: ranking and thresholding. The ranking induced by $\rho$-values closely resembles that of Lfdr-based methods, while the thresholding step aligns with conventional p-value procedures. Therefore, the proposed framework guarantees FDR control under mild assumptions; it maintains the integrity of the structural information encoded by the summary statistics and the auxiliary covariates, and hence can be asymptotically optimal. We demonstrate the advantages of the $\rho$-value framework through comprehensive simulations and analyses of two real datasets: one from microbiome research and another related to attention deficit hyperactivity disorder (ADHD).
\end{abstract}


  \section{Introduction}

With the advent of big data and increased data availability, multiple testing has become an increasingly critical challenge in modern scientific research. For instance, in microbiome-wide association studies (MWAS), investigating the relationship between microbiome features and complex host traits typically involves testing thousands of variables across numerous microorganisms. Without proper correction for multiplicity, such analyses are prone to inflated false positive rates.
Similarly, in magnetic resonance imaging (MRI) studies aimed at identifying functional regions of the human brain for clinical diagnosis or medical research, the massive volume of high-resolution 3D imaging data complicates the task of simultaneous inference. High-dimensional regression settings—such as those encountered in gene association studies—present another example, where thousands of genes are tested for associations with drug sensitivities, particularly in cancer research.

A widely adopted measure of Type I error in multiple testing is the false discovery rate (FDR; \citet{benjamini1995controlling}), defined as the expected proportion of false positives among all discoveries. Since its introduction, FDR has rapidly become a central concept in modern statistics and a primary tool for large-scale inference across a wide range of scientific disciplines.
At a high level, most FDR-controlling procedures that control FDR  operate in two steps: first rank all hypotheses according to some significance indices and then reject those with index values less than or equal to some threshold. 
In this paper, we propose a new multiple testing framework built upon a novel concept called $\rho$-values. This framework unifies the commonly used p-value and local false discovery rate (Lfdr) approaches while offering several advantages over both. Moreover, the $\rho$-value framework is closely connected to e-value based methods, providing greater flexibility in handling data dependencies.
   In what follows, we begin by reviewing conventional multiple testing practices and identifying their limitations, and then introduce the proposed $\rho$-value framework, highlighting its theoretical and practical contributions.


Several popular FDR-controlling procedures use p-values as significance indices for ranking hypotheses \citep[e.g.,][]{benjamini1995controlling,genovese2006false,liu2013gaussian,lei2018adapt,cai2022laws}. Typically, p-values are derived by applying a probability integral transform to well-known test statistics. 
  For example, \cite{li2019multiple} uses a permutation test, \cite{roquain2009optimal} employs a   Mann--Whitney U test and \cite{cai2022laws} adopts a t-test.
  However, the p-value based methods can be inefficient because the conventional p-values do not  incorporate information from the alternative distributions  \citep[e.g.,][]{sun2007oracle,ZAP}.  The celebrated Neyman--Pearson lemma states that the optimal statistic for testing a single  hypothesis is the likelihood ratio. In the multiple testing context, the local false discovery rate (Lfdr) serves as the natural analog of the likelihood ratio statistic \citep[e.g.,][]{efron2001empirical,efron2003robbins,aubert2004determination,hong2009local,sarkar2017local}. 
    It has been shown that a ranking and thresholding procedure based on Lfdr is asymptotically optimal for FDR control \citep{sun2007oracle,xie2011optimal,tony2019covariate}. Subsequently, \cite{heller2021optimal} demonstrates that such Lfdr-based procedures are in fact exact optimal among all FDR-controlling rules.
 Nevertheless, the performance of Lfdr-based methods critically depends on the accurate estimation of the Lfdr, which itself requires integrating information across all test statistics—a task that can be quite challenging in practice \citep{marandon2024adaptive}. This challenge becomes even more pronounced when incorporating side information, a common necessity in real-world applications. For example, in the MWAS dataset analyzed in Section \ref{sec51}, the proportion of zeros across samples for each operational taxonomic unit (OTU) serves as side information, capturing both biological and technical variability \citep{xia2020correlation}. Similarly, in the MRI dataset analyzed in Section \ref{sec52}, spatial coordinate indices act as side information, enabling the use of spatial structures to enhance signal detection and interpretability \citep{paloyelis2007functional}.
   To overcome the Lfdr estimation challenge, several methods have been developed that aim to approximate Lfdr-based procedures using weighted p-values \citep[e.g.,][]{lei2018adapt,li2019multiple,liang2023lasla}. While promising, these Lfdr-mimicking approaches often rely on strong model assumptions or remain suboptimal in practice.
   Recent developments in conformal inference offer an appealing alternative by constructing provably valid marginal p-values without requiring explicit knowledge of the null distribution. These conformal p-values use data-driven calibration and provide finite-sample marginal FDR guarantees under relatively mild assumptions \citep{bates2023testing,marandon2024adaptive}. Nevertheless, this added flexibility comes at the cost of weaker FDR guarantees compared to methods that leverage exact null information.

  
  To address the aforementioned challenges, this article introduces a novel concept—the $\rho$-value—which adopts the form of a likelihood ratio while offering greater flexibility in the choice of density functions compared to the traditional Lfdr. Building on this concept, we propose a new and flexible multiple testing framework that unifies p-value-based and Lfdr-based approaches.
The $\rho$-value-based Benjamini--Hochberg (BH) procedure, including its weighted variant (analogous to weighted p-values), also follows the standard two-step structure: first, all hypotheses are ranked according to their (weighted) $\rho$-values, with the ranking designed to approximate that of the Lfdr; second, hypotheses with (weighted) $\rho$-values less than or equal to a data-driven threshold are rejected. The thresholding strategy is analogous to that used in conventional p-value procedures, thereby ensuring both interpretability and control of the FDR.

  Compared to existing frameworks, the proposed $\rho$-value framework offers several notable advantages.
First, when carefully constructed, $\rho$-values produce a ranking of hypotheses that coincides with that of Lfdr statistics. As a result, procedures based on $\rho$-values can achieve asymptotic optimality.
Second, the thresholding strategy in $\rho$-value-based methods mirrors that of p-value-based procedures. This similarity allows the proposed methods to inherit desirable theoretical properties of p-value-based approaches. Importantly, FDR control in the $\rho$-value framework does not rely on consistent estimation of Lfdr statistics, making the approach significantly more flexible than traditional Lfdr-based methods.
Third, side information can be seamlessly incorporated into $\rho$-value procedures through a simple weighting scheme, enhancing the ranking of hypotheses and thereby improving power. This integration is often more straightforward than in the Lfdr framework, where side information complicates density estimation.
Fourth, our proposed $\rho$-BH procedure demonstrates superior power compared to the e-BH procedure, and the weighted $\rho$-BH procedure consistently outperforms the weighted BH procedure in terms of detection power.
Finally, the proposed framework provides a unified perspective that bridges p-value-based and Lfdr-based methodologies. In particular, we show that these two paradigms are more closely related than previously suggested \citep[e.g.,][]{sun2007oracle,ZAP}. 

The paper is structured as follows. Section \ref{sec:method} starts with the problem formulation. It then introduces the $\rho$-BH procedure and its variations. Sections \ref{sec:simu} and \ref{sec:real} present numerical  comparisons of the proposed methods and other competing approaches using simulated and real  data, respectively. More discussions of the proposed framework are provided in Section \ref  {sec:discuss}. The weighted $\rho$-BH procedures and the corresponding theories, as well as all technical proofs are collected in the Appendix.

	\section{Methodology}\label{sec:method}
	In this section, we begin by introducing the problem formulation and the motivation behind the development of $\rho$-value-based procedures. Subsequently, in Sections \ref{sec:rhobh}–\ref{sec:side}, we present a series of $\rho$-value-based multiple testing methods. Specifically, we introduce both oracle and data-driven versions of the $\rho$-BH procedure, followed by extensions that incorporate side information into the $\rho$-value framework.
\subsection{Problem formulation}\label{sec:prob}
Suppose our goal is to simultaneously test the following $m$ hypotheses: 
	\begin{equation*}
		\label{eq:model1}
		H_{0, i}: \theta_i=0 \quad \text{versus} \quad  H_{1, i}: \theta_i=1,~i=1,\ldots,m.
  \end{equation*}
  Assume that we observe independent summary statistics $\{X_i\}_{i=1}^m$  arising from the following random mixture model: 
	\begin{equation}\label{eq:prob}
		\theta_i\overset{iid}{\sim }\text{Ber}(\pi),\quad X_i|\theta_i\overset{ind}{\sim} (1-\theta_i)f_0+\theta_i f_1,
	\end{equation}
	where $\pi = P(\theta_i=1)$ and $f_0$ and $f_1$ respectively represent the null and alternative density functions of $X_i$.
	 By convention, the null density $f_0$ is assumed to be known.
	Such a model has been widely adopted in many large-scale inference problems \citep[e.g.,][]{efron2004large,efron2007testing,jin2007estimating}. 
	For simplicity, we assume homogeneous $\pi$ and $f_1$ in Model \eqref{eq:prob} for now, and it will be extended to heterogeneous scenarios in later sections. 
	In addition,  Model \eqref{eq:prob} does not necessarily correspond to the actual data generation process. Instead, it only serves as a hypothetical model to motivate our methodology.
	
	Denote by $\bm\delta=(\delta_1, \cdots, \delta_m)\in\{0,1\}^m$ an $m$-dimensional decision vector, where $\delta_i=1$ means we reject $H_{0,i}$, and $\delta_i=0$ otherwise. In large-scale multiple testing problems, false positives are inevitable if one wishes to discover non-nulls with a reasonable power. Instead of aiming to avoid any false positives, \cite{benjamini1995controlling} introduces the FDR, i.e., the expected proportion of false positives among all selections, written formally as
	$\text{FDR}(\bm \delta) =  E\left[\{{\sum_{i=1}^{m}(1-\theta_i)\delta_i}\}/\{\max\{\sum_{i=1}^{m}\delta_i, 1\}\}\right],$
	and a practical goal is to control the FDR at a pre-specified significance level.
	A closely related quantity of FDR is the marginal false discovery rate (mFDR), defined by $\text{mFDR}(\bm \delta) ={{E}\{\sum_{i=1}^{m}(1-\theta_i)\delta_i\}}/{{E}(\sum_{i=1}^{m}\delta_i)}.$
	Under certain first- and second-order conditions on the number of rejections, the mFDR and the FDR are asymptotically equivalent \citep{genovese2002operating,basu2018weighted,tony2019covariate}. The mFDR criterion is often employed to facilitate methodological development and derive optimality results in large-scale testing problems.
	An ideal multiple testing procedure should both control the FDR (or mFDR) at a pre-specified nominal level $\alpha$ and maximize statistical power, which is quantified by the expected number of true positives:
$\mbox{ETP}(\bm \delta)= E\left(\sum_{i=1}^{m} \theta_i \delta_i\right).$
	We call a multiple testing procedure \textit{valid} if {it controls the mFDR asymptotically at the nominal level $\alpha$}, and \textit{optimal} if it has the largest ETP among all valid procedures. We call $\pmb{\delta}$ asymptotically optimal  if $\mbox{ETP}(\bm \delta)/	\mbox{ETP}(\bm \delta')\geq1+o(1)$ for any decision rule $\pmb{\delta}'$ that controls mFDR at the pre-specified level $\alpha$ asymptotically.

	\subsection{Motivation, $\rho$-value and the $\rho$-BH procedure}\label{sec:rhobh}

The classical BH procedure \citep{benjamini1995controlling} remains one of the most widely used approaches for multiple testing. It defines the decision rule as $\pmb{\delta}_{BH}=(\mathbb{I}(p_1\leq p_{(k)}),\ldots, \mathbb{I}(p_m\leq p_{(k)}))$, where $p_i$ is the p-value for $H_{0,i}$ and $k=\max\{ i:m p_{(i)}\leq \alpha i\}$. 
It is demonstrated in \citep{benjamini1995controlling} that the BH procedure in fact controls FDR at level $(1-\pi)\alpha$. Therefore, a straightforward refinement of BH procedure, which we refer to as the adjusted BH procedure, selects $k=\max\{ i:m(1-\pi) p_{(i)}\leq \alpha i\}$. 
Intuitively, the adjusted BH procedure ranks hypotheses by their p-values and selects a threshold $p_{(k)}$ such that the estimated FDR, $(1-\pi)mp_{(k)}/k$, does not exceed $\alpha$. 
Though the adjusted BH procedure guarantees FDR control under independence, p-values generally lack information about the alternative distribution, which may result in suboptimal power.

An alternative approach is based on the Lfdr \citep{efron2004large}, defined as $$\mbox{Lfdr}_i\equiv P(\theta_i=0|X_i)=\dfrac{(1-\pi)f_0(X_i)}{(1-\pi)f_0(X_i)+\pi f_1(X_i)}. $$
Let $\text{Lfdr}_{(1)}\leq \ldots\leq \text{Lfdr}_{(m)}$ be the order statistics of $\text{Lfdr}_1,\ldots, \text{Lfdr}_m$.
It is shown in \cite{sun2007oracle} that the decision rule $\pmb{\delta}_{SC}=(\mathbb{I}(\text{Lfdr}_1\leq \text{Lfdr}_{(k)}),\ldots, \mathbb{I}(\text{Lfdr}_m\leq \text{Lfdr}_{(k)}))$, where $k =\max \{i:i^{-1}  \sum_{j=1}^{i}\text{Lfdr}_{(j)} \leq \alpha\}$,
is asymptotically optimal among all mFDR control rules.  We refer to this rule as the SC procedure.
The intuition behind the SC procedure is twofold: first, the Lfdr statistics provide an optimal ranking of hypotheses based on their likelihood of being null; second, since the mFDR is an increasing function of the threshold, the threshold $\text{Lfdr}_{(k)}$ is chosen such that the estimate of FDR, given by $k^{-1} \sum_{j=1}^{k}\text{Lfdr}_{(j)} $,   is just below $\alpha$.
Therefore, for the SC procedure, consistent estimates of the $\text{Lfdr}_i$'s are essential to ensure that $k^{-1} \sum_{j=1}^{k}\text{Lfdr}_{(j)} $ approximates the true mFDR when the threshold is set at $\text{Lfdr}_{(k)}$.

Our proposal aims to combine the strengths of the SC and the adjusted BH procedures. To motivate this, consider the idealized setting where $f_1$ is known. Define $\rho_i\equiv f_0(X_i)/f_1(X_i)$. Since $f_0$ is known, the null cumulative distribution function for $\rho_i$ is also known; we denote it by $c(\cdot)$. By definition, $c(\rho_i)$ is a valid p-value for testing $H_{0,i}$. Consequently, we can apply the adjusted BH procedure directly to the transformed values $(c(\rho_1), \ldots, c(\rho_m))$, yielding a procedure we denote by $\pmb{\delta}_\rho$. 

The new procedure $\pmb{\delta}_\rho$ can be conceptualized in two steps. First, ranking null hypotheses according to $c(\rho_i)$. Second, thresholding using the adjusted BH rule. 
Since $c(\cdot)$ is an increasing function, it is clear that ranking by $c(\cdot)$ is equivalent to ranking by $\text{Lfdr}_i$. Moreover, the threshold chosen by the adjusted BH procedure is sharp in the sense that increasing the threshold to $c(\rho_{(k+1)})$, where $k=\max\{ i:m(1-\pi) c(\rho_{(i)})\leq \alpha i\}$,  would violate the FDR control constraint at level $\alpha$. Thus, it is intuitively clear that $\pmb{\delta}_\rho$ is equivalent to $\pmb{\delta}_{SC}$ asymptotically. The power equivalence is formalized in the next theorem, while the FDR validity for the general $\rho$-value defined in Definition \ref{def:rho} below is deferred to Theorem~\ref{thm1}.

	\begin{theorem}\label{thm:opt1}
	Let $\rho_i=f_0(X_i)/f_1(X_i)$ and suppose $X_i$'s are independent. Let $\pmb{\delta}$ be any testing rule with {$\mbox{mFDR}(\pmb{\delta})\leq \alpha$} {asymptotically}. Suppose
	\begin{enumerate}[label=(A\arabic*), series=A, itemindent=10pt]	
		\item \label{A1} 
		$m P\left(\rho_i \leq  \frac{\alpha\pi}{(1-\pi)(1-\alpha)} \right) \rightarrow \infty $.
	\end{enumerate}
	Then we have $\mbox{ETP}(\pmb{\delta}_\rho)/\mbox{ETP}(\pmb{\delta})\geq 1+o(1)$. 
\end{theorem}
	\begin{remark}\label{remark1}
If the SC procedure results in at least one rejection with probability tending to $1$, it implies that $m P(\text{Lfdr}_i\leq\alpha)\rightarrow \infty$ as $m\rightarrow\infty$.
	This serves as an equivalent condition for Assumption \ref{A1}, which thus represents a mild condition.
	
\end{remark}

At first glance, the procedure $\pmb{\delta}_\rho$ may appear to offer no new advantages over the SC procedure. However, a crucial distinction lies in its {\it robustness}: the validity of $\pmb{\delta}_\rho$ remains intact even if the alternative density $f_1(\cdot)$ is replaced by an arbitrary function $g(\cdot)$. That is, even when $\rho_i = f_0(X_i)/g(X_i)$ is computed using a misspecified or surrogate alternative, the resulting adjusted BH procedure applying to p-values $c(\rho_i)$ still controls the FDR under independence.

This robustness implies that $\bm{\delta}_\rho$ can be viewed as a generalization and improvement of the SC procedure. 
On one hand, it retains FDR control under independence without requiring consistent estimation of the true Lfdr or the true $f_1$. On the other hand, when $\pi$ and $\text{Lfdr}_i$ are known, the  procedure $\bm{\delta}_\rho$ is asymptotically optimal.

Generally, we refer to $\rho_i = f_0(X_i)/g(X_i)$, for any density function $g(\cdot)$, as a $\rho$-value of $X_i$, and the corresponding procedure $\bm{\delta}_\rho$ as the $\rho$-BH procedure.
	\begin{definition}\label{def:rho}
		Suppose $X$ is a summary statistic and $f_0(\cdot)$ is the density of $X$ under the null. A $\rho$-value of $X$ is defined as $$\rho\equiv f_0(X)/g(X),$$
		where $g(\cdot)$ is any density function satisfying $g(X) \neq 0$.
	\end{definition}
A summary of the $\rho$-BH algorithm is provided in Algorithm~\ref{alg1}, and its theoretical validity is established in Theorem~\ref{thm1}.

\begin{algorithm}
	\caption{The $\rho$-BH procedure} \label{alg1}
	\begin{description}
		\item[Input:] $\{X_i\}_{i=1}^m$; a predetermined density function $g(\cdot)$; non-null proportion $\pi$; desired FDR level $\alpha$.
		\item[1.] Calculate the $\rho$-values $\rho_i = f_0(X_i)/g(X_i)$, for $i=1,\ldots,m$.
		\item[2.] Sort the $\rho$-values from smallest to largest  $\rho_{(1)} \leq \cdots \leq \rho_{(m)}$.
		\item[3.] Compute the null distribution function of $\rho_i$'s, and denote it by $c(\cdot)$.
		\item[4.] Let $k = \max_{1 \leq j \leq m} \left[c(\rho_{(j)})\leq (\alpha j)/\{m(1-\pi)\} \right]$. 
		\item[Output:] The rejection set \{$i=1,\ldots,m$: $\rho_i \leq \rho_{(k)}$\}.
	\end{description}
\end{algorithm}

	\begin{theorem}\label{thm1}
		Assume that the null $\rho$-values are mutually independent and are independent of the non-null $\rho$-values, then $\text{FDR}_{\text{Algorithm } \ref{alg1}} \leq \alpha$. 
	\end{theorem}

\subsection{The data-driven $\rho$-BH procedure}\label{sec:dd}
In practice, $f_1$ and $\pi$ are usually unknown and need to be estimated from the data. The problem of estimating non-null proportion has been discussed extensively in the literatures \citep[e.g.,][]{storey2002direct, meinshausen2006estimating, jin2007estimating,chen2019uniformly}. To ensure valid mFDR control, we require the estimator $\hat{\pi}$ to be conservative consistent, defined as follows.
\begin{definition}
	\label{def:unifcon}
	An estimator $\hat{\pi}$ is a conservative consistent estimator of $\pi$, if $|\hat{\pi} - \tilde{\pi}|\overset{P}{\rightarrow} 0$ as $m\rightarrow \infty$, for some $\tilde{\pi}$ satisfying $0\leq\tilde{\pi}\leq\pi$.
\end{definition}

One possible choice of such $\hat{\pi}$ is the Storey estimator as provided by the following proposition.
\begin{proposition}\label{prop1}
	The estimator 
	$
	\hat{\pi}^\tau=1-{\#\{i:c(\rho_i)\geq \tau\}}/\{m(1-\tau)\}
	$
	proposed in \cite{storey2002direct}
	is conservative consistent for any $\tau$ satisfying $0\leq \tau\leq 1$.
\end{proposition}

The problem of estimating $f_1$ is more complicated. 
If we use the entire sample $\{X_i\}_{i=1}^m$ to construct $\hat{f}_1$ and let $\rho_i=f_0(X_i)/\hat{f}_1(X_i)$, then $\rho_i$'s are no longer independent even if $X_i$'s are. One possible strategy to circumvent this dependence problem is to use sample splitting.
More specifically, we can randomly split the data into two disjoint halves and use the first half of the data to estimate the alternative density for the second half, i.e., $\hat{f}_1^{\scriptscriptstyle (2)}$ (e.g., we can use the estimator proposed in \cite{HART}),  then the $\rho$-values for the second half can be calculated by $f_0(X_i)/\hat{f}_1^{\scriptscriptstyle (2)}(X_i)$.  Hence, when testing the second half of the data, $\hat{f}_1^{\scriptscriptstyle (2)}$ can be regarded as predetermined and independent of the data being tested. 	The decisions on the first half of the data can be obtained by switching the roles of the first and the second halves and repeating the above steps. If the FDR is controlled at level $\alpha$ for each half, then the overall mFDR is also controlled at level $\alpha$ asymptotically.
We summarize the above discussions in Algorithm \ref{alg2}.

\begin{algorithm}
	\caption{The data-driven $\rho$-BH procedure}\label{alg2}
	\begin{description}
		\item[Input:] $\{X_i\}_{i=1}^m$; desired FDR level $\alpha$.
		\item[1.] Randomly split the data into two disjoint halves
		$
		\{X_i\}^m_{i=1}=\{X_{1,i}\}^{m_1}_{i=1}\cup\{X_{2,i}\}^{m_2}_{i=1}
		$, where $m_1=\lfloor m/2 \rfloor$. 
		\item[2.] Use $\{X_{1,i}\}^{m_1}_{i=1}$ to construct the second half alternative estimate {$\hat{f}_1^{\scriptscriptstyle (2)}$} and a conservative consistent estimate $\hat{\pi}_2$.
		\item[3.] Run Algorithm \ref{alg1} with $\{X_{2,i}\}_{i=1}^{m_2}$, $\hat{f}_1^{\scriptscriptstyle (2)}$, $\hat{\pi}_2$, $\alpha$ as inputs.
		\item[4.] Switch the roles of  $\{X_{1,i}\}^{m_1}_{i=1}$ and $\{X_{2,i}\}^{m_2}_{i=1}$. Repeat Steps 2 and 3, and combine rejections.
		\item[Output:] The combined rejection set.
	\end{description}
\end{algorithm}

\begin{remark}\label{remark2}
	{A natural question for the data-splitting approach is whether it will negatively impact the power. 
		Suppose that $\hat{f}_1^{\scriptscriptstyle (1)}$, $\hat{f}_1^{\scriptscriptstyle (2)}$ are consistent estimators for some function $g$, and $\hat{\pi}_1$, $\hat{\pi}_2$ are consistent estimators for some constant $\tilde \pi$. Denote by $t_\alpha$ the threshold selected by Algorithm \ref{alg1} with $g$ and $\tilde \pi$ as inputs, on full data. Then it is expected that the thresholds $\hat t_1$ and $\hat t_2$ selected by Algorithm \ref{alg2} for each half of the data will both converge to $t_\alpha$, provided that the empirical distributions of the two halves are similar. 
		Therefore, the decision rule by data-splitting tends to be as powerful as the rule based on the full data.}
\end{remark}


Next we provide the theoretical guarantee for Algorithm \ref{alg2} in the following theorem.
\begin{theorem}\label{thm2}
	Assume that $X_i$'s are independent.  
	Denote by $\{\hat{\rho}_{d,i}\}_{i=1}^{m_d}$, $\hat{\rho}_{d,(k_d)}$ and $\hat{\pi}_d$ the $\rho$-values, selected thresholds and the estimated alternative proportions obtained from Algorithm \ref{alg2}, for the first and second halves of the data respectively, $d=1,2$.
	Denote by $\hat{c}_d$ the null distribution function for $\hat{\rho}_{d,i}$. 
	Suppose $\hat{\pi}_d >0$ and $|\hat{\pi}_d - \tilde{\pi}_d|\overset{P}{\rightarrow} 0$ for some $\tilde{\pi}_d$ satisfying $0\leq\tilde{\pi}_d\leq\pi$,
	and let $\tilde Q_d(t) = {(1 - \tilde\pi_d)\hat c_{d}(t)}/{P(\hat \rho_{d,i} \leq t)}$ and $t_{d,L} = \sup\{t>0: \tilde Q_d(t)\leq\alpha\}$, $d=1,2$.
	Assume the following hold
	\begin{enumerate}[resume*=A]
		\item \label{A2}  $\hat{\rho}_{d,(k_d)} \geq \nu\frac{\hat{\pi}_d}{1-\hat{\pi}_d}$ and  $P(\hat \rho_{d,i} \leq \nu\frac{\hat{\pi}_d}{1-\hat{\pi}_d}) >c$, for some constants $\nu, c>0$;
		\item \label{A3} $\limsup_{t\rightarrow 0^+}\tilde Q_d(t)<\alpha$, $\liminf_{t\rightarrow \infty}\tilde Q_d(t)>\alpha$;
		\item \label{A4}  $\inf_{t \geq t_{d,L} + \epsilon_t}\tilde Q_d(t) \geq \alpha + \epsilon_\alpha$, and $\tilde Q_d(t)$ is strictly increasing in $t\in(t_{d,L} - \epsilon_t, t_{d,L} + \epsilon_t)$, for some constants $\epsilon_\alpha, \epsilon_t > 0$.
	\end{enumerate}
	Then we have $\lim_{m\rightarrow \infty}\text{mFDR}_{\text{Algorithm } \ref{alg2}} \leq \alpha$.
\end{theorem}

\begin{remark}		
	Theorem \ref{thm:opt1} and the oracle rule in \cite{sun2007oracle} imply that, when the alternative density and the non-null proportion are estimated by the truths, the threshold of the $\rho$-values should be at least { $\alpha\pi/\{(1-\pi)(1-\alpha)\}$}. Since $\hat{\pi}_d$'s are conservative consistent, we have { $\hat{\pi}_d/(1-\hat{\pi}_d)$} converges in probability to a number less than { $\pi/(1-\pi)$}. Therefore, the first part of \ref{A2} is mild. Moreover, by setting $\nu$ equal to some fixed number, say {\ $\alpha/(1-\alpha)$}, the first part of \ref{A2} can be easily checked {numerically}. 
	The second part of \ref{A2} is only slightly stronger than the condition that the total number of rejections for each half of the data is of order $m$. 
	It is a sufficient condition to show that the estimated FDP, { $m_d(1-\hat{\pi}_d)\hat{c}_d(t)/\sum_{i=1}^{m_d}\mathbb{I}(\hat{\rho}_{d,i}\leq t)$}, is close to $\tilde{Q}_d(t)$, and it can be easily relaxed if $\hat\pi$ satisfies certain convergence rate. 
	\ref{A3} is also a reasonable condition, it excludes the trivial cases where no null hypothesis can be rejected or all null hypotheses are rejected. If $\tilde{Q}_d$'s are continuous, then the first part of \ref{A4} is implied by \ref{A3} and the definition of $t_{d,L}$. The second part of \ref{A4} can be easily verified numerically and it is also mild under the continuity of $\tilde{Q}_d$.
	Finally, all of the above conditions are automatically satisfied in the oracle case where $\pi$ and $\text{Lfdr}_i$ are known.
\end{remark}

	\subsection{$\rho$-BH under dependence}\label{sec:dep}

So far, we have assumed that the summary statistics $X_i$'s are independent. However, in many applications, the $X_i$'s are observed across related groups, spatial locations, or time points, leading to inherent dependencies among the observations. To account for arbitrary dependence structures, \citet{benjamini2001control} proposed a conservative adjustment to the BH procedure. Specifically, they showed that the adjusted BH procedure with nominal levell $\alpha$ controls the FDR at level $\alpha S(m)$, where 
$
S(m) = \sum_{i=1}^m \frac{1}{i} \approx \log m
$
is known as the Benjamini--Yekutieli (BY) correction factor.
Accordingly, we obtain the following result for Algorithm~\ref{alg1} under the BY correction.

\begin{theorem}\label{thm:by}
	By setting the target FDR level to $\alpha / S(m)$, we have
	$
	\text{FDR}_{\text{Algorithm } \ref{alg1}} \leq \alpha
	$
	under arbitrary dependence.
\end{theorem}

An alternative approach for handling arbitrary dependence involves the use of e-values \citep{wang2022false,vovk2021values}. A non-negative random variable $e$ is called an e-value if $\mathbb{E}(e) \leq 1$, where the expectation is taken under the null hypothesis. Using Markov's inequality, it can be shown that the reciprocal of an e-value is a valid p-value \citep{wang2022false}. Since
$$
\mathbb{E}(1/\rho) = \int \frac{g(x)}{f_0(x)} f_0(x) dx = \int g(x) dx = 1,
$$
it follows that $1/\rho$ is an e-value. It is further shown in \citep{wang2022false} that if the reciprocals of e-values are used as input for the BH procedure, the resulting method—known as the e-BH procedure—controls the FDR at the target level under arbitrary dependence.

If one opts for either the BY correction or the e-BH procedure, then the data-splitting step in data-driven procedures such as Algorithm~\ref{alg2} is no longer necessary.

We further note that our earlier FDR control results under independence can be extended to settings with weakly dependent $\rho$-values. Specifically, the technical tools developed in \citep{liu2013gaussian, chen2018, cai2022laws} can be employed to establish asymptotic control of error rates under weak dependence. However, to maintain focus on the introduction of the proposed $\rho$-value-based testing framework, we defer detailed discussions of weak dependence to future work.




	\subsection{$\rho$-BH with side information}\label{sec:side}
	In many scientific applications, additional covariate information—such as patterns of the signals—is often available. Consequently, the problem of multiple testing with side information has garnered significant attention and has emerged as an active area of research in recent years \citep[e.g.,][]{du2014single,lei2018adapt,ramdas2019unified,li2019multiple,ignatiadis2021covariate,cao2022optimal,zhang2022covariate,liang2023lasla}. As demonstrated in these studies, appropriately incorporating such side information can substantially improve both the power and the interpretability of simultaneous inference procedures.
	Let $X_i$ denote the primary statistic and $s_i\in \mathbb{R}^l$ the side information.
	Upon observing $\{(X_i,s_i)\}_{i=1}^m$, we test
	$H_{0,i}:\theta_i=0$ versus $H_{1,i}:\theta_i=1$ for $i=1,\ldots,m$. 
{To motivate our analysis, we}
model the data generation process as follows.
	\begin{align}\label{side}
	\theta_i|s_i\overset{ind}{\sim }\text{Ber}(\pi(s_i)), \quad X_i|s_i,\theta_i\overset{ind}{\sim} (1-\theta_i)f_0(\cdot|s_i)+\theta_i f_1(\cdot|s_i),\quad i=1,\ldots,m.
	\end{align}
{Again, similarly as \cite{tony2019covariate,liang2023lasla}, we do not assume that the data are generated exactly as described in Model \eqref{side}. Such model only serves as a tool to motivate methodological development.}
 As before, we assume the null distributions $f_0(\cdot|s_i)$ are known. While this assumption may appear strong at first glance, in practice the null distribution frequently does not depend on the auxiliary or side information variables \citep{tony2019covariate, ZAP}.
 The $\rho-$value framework can be easily  adapted to incorporate such side information by introducing a weighting scheme that leverages the additional side information associated with each hypothesis.
  We define the $\rho$-values by $\rho_i={f_0(X_i|s_i)}/{g(X_i|s_i)}$ for some density function $g(\cdot|s_i)$. 
	Let $\eta:\mathbb{R}^l\rightarrow (0,1)$ denote a predetermined function, and $c_i(\cdot)$ represent the null distribution of $\rho_i$. 
	Then we incorporate the side information through a $\rho$-value weighting scheme 
	by choosing an appropriate function $\eta(\cdot)$ to determine the weights. The detailed steps are summarized in Algorithm \ref{alg4}.
	
	\begin{algorithm}[ht]
		\caption{The $\rho$-BH procedure with side information}\label{alg4}
		\begin{description}
			\item[Input:]  $\{X_i\}_{i=1}^m$; $\{s_i\}_{i=1}^m$; predetermined density functions $\{g(\cdot|s_i)\}_{i=1}^m$; non-null proportions $\{\pi(s_i)\}_{i=1}^m$; predetermined $\{\eta(s_i)\}_{i=1}^m$; desired FDR level $\alpha$.
			\item[1. for $i=1$ to $m$ do:]
			\item[] \qquad Calculate the $\rho$-values $\rho_i = f_0(X_i|s_i)/g(X_i|s_i)$.
			\item[] \qquad Compute the null distribution functions of each $\rho_i$, and denote them by $\{c_i(\cdot)\}_{i=1}^m$.
			\item[] \qquad Let $w_i=\frac{\eta(s_i)}{1-\eta(s_i)}$, and compute the weighted $\rho$-values $q_i=\rho_i/w_i$.
			\item[\quad \ end for] 
			\item[2.] Sort the weighted $\rho$-values from smallest to largest  $q_{(1)} \leq \cdots \leq q_{(m)}$.
			\item[3.] Let $k = \text{max}_{1 \leq j \leq m} \left\{\sum_{i=1}^{m} \{1-\pi(s_i)\}c_i(q_{(j)}w_i)\leq \alpha j \right\} $.
			\item[Output:] The rejection set $\{i=1,\ldots,m: q_i\leq q_{(k)}\}$.
		\end{description}
	\end{algorithm}

	In contrast to the $\rho$-BH procedure,  Algorithm \ref{alg4} is no longer equivalent to the adjusted BH procedure with $\{c_i(\rho_i)/ w_i\}$'s as the inputs since the rankings produced by $\{c_i(\rho_i) /w_i\}$'s and $\{\rho_i /w_i\}$'s are different.
	The ideal choice of $g(\cdot|s_i)$ is again $f_1(\cdot|s_i)$, while the ideal choice of $\eta(\cdot)$ is $\pi(\cdot)$, and the rationale is provided as follows.
	Define the conditional local false discovery rate (Clfdr, \cite{HART,tony2019covariate}) as $$\mbox{Clfdr}_i\equiv\frac{\{1-\pi(s_i)\}f_0(X_i|s_i)}{\{1-\pi(s_i)\}f_0(X_i|s_i)+\pi(s_i)f_1(X_i|s_i)  }.$$
	\cite{tony2019covariate} shows that a ranking and thresholding procedure based on Clfdr is asymptotically optimal for FDR control. If we take $g(\cdot|s_i)$ to be $f_1(\cdot|s_i)$ and $\eta(\cdot)$ to be $\pi(\cdot)$, then the ranking produced by $\rho_i/w_i$'s is identical to that produced by Clfdr statistics. 
	However, the validity of the data-driven methods proposed in \cite{tony2019covariate} and \cite{HART} relies on the consistent estimation of Clfdr$_i$'s. In many real applications, 
	it is extremely difficult to accurately estimate Clfdr even when the dimension of $s_i$ is moderate \citep{cai2022laws}. 
	In contrast, the mFDR guarantee of Algorithm \ref{alg4} does not rely on any of such Clfdr consistency results and our proposal is valid under much weaker conditions as demonstrated by the next theorem.

\begin{theorem}\label{thm4}
	Assume that $\{X_i, \theta_i\}_{i=1}^m$ are independent. Let $Q(t) = \frac{\sum_{i=1}^m \{1 - \pi(s_i)\} c_i(w_i t)}{E\{\sum_{i=1}^m \mathbb{I}(q_i \leq t)\}}$ and $t_L = \sup\{t>0: Q(t)\leq\alpha\}.$ Based on the notations from Algorithm \ref{alg4} and suppose
	\begin{enumerate}[resume* = A]
		\item \label{A5}  $q_{(k)} \geq \nu$ and $\sum_{i=1}^m P(q_i \leq \nu) \rightarrow \infty$ as $m\rightarrow\infty$, for some $\nu > 0$;
		\item \label{A6}  $\limsup_{t\rightarrow 0^+}Q(t)<\alpha$; $\liminf_{t\rightarrow \infty}Q(t)>\alpha$;
		\item \label{A7}  $\inf_{t \geq t_L + \epsilon_t}Q(t) \geq \alpha + \epsilon_\alpha$, $Q(t)$ is strictly increasing in $t\in(t_L - \epsilon_t, t_L + \epsilon_t)$, for some constants $\epsilon_\alpha, \epsilon_t > 0$.
	\end{enumerate}
	Then we have
$
		\lim_{m\rightarrow\infty} \text{mFDR}_{\text{Algorithm } \ref{alg4}} \leq \alpha.
$
\end{theorem}

	{The validity of the above theorem allows flexible choices of functions $g(\cdot|s_i)$ and the weights $w_i$. Hence, similarly as the comparison between $\rho$-value and Lfdr, the $\rho$-value framework with side information is again much more flexible than the Clfdr framework that requires the consistent estimation of the Clfdr statistics.}
	
	We also remark that \cite{cai2022laws} recommends using ${\pi(s_i)}/{1 - \pi(s_i)}$ as weights for p-values derived from two-sample t-statistics. However, their justification is primarily heuristic, and the advantage of using ${\pi(s_i)}/{1 - \pi(s_i)}$ over the alternative ${1}/{1 - \pi(s_i)}$ is not rigorously established. In contrast, we present the following optimality result, which offers a more principled justification for the use of ${\pi(s_i)}/{1 - \pi(s_i)}$.

	\begin{theorem}\label{thm:opt2}
		Assume that $\{X_i, \theta_i\}_{i=1}^m$ are independent.
		Denote by $\pmb{\delta}_\rho$ the rule described in Algorithm \ref{alg4} with $\eta(\cdot)=\pi(\cdot)$ and $g(\cdot|s_i)=f_1(\cdot|s_i)$, and let $\pmb{\delta}$ be any other rule that controls mFDR at level $\alpha$ asymptotically. Based on the notations from Algorithm \ref{alg4} and suppose 
		\begin{enumerate}[resume* = A]
		\item \label{A8}  $\sum_{i=1}^m P(q_i \leq \frac{\alpha}{1-\alpha}) \rightarrow \infty$ as $m\rightarrow\infty$.
		\end{enumerate}
Then we have $\mbox{ETP}(\pmb{\delta}_\rho)/\mbox{ETP}(\pmb{\delta})\geq 1+o(1)$. 
	\end{theorem}
	\begin{remark}
	Assumptions \ref{A5}-\ref{A7} are automatically satisfied under the conditions assumed by Theorem \ref{thm:opt2}. Therefore, in such ideal setting, Algorithm \ref{alg4} is  optimal among all testing rules that asymptotically control mFDR at level $\alpha$.
	In addition, Theorem \ref{thm:opt2} implies that the weighted BH procedure \citep{genovese2006false} based on the ranking of $\{c_i(\rho_i)/w_i\}$ is suboptimal.
	\end{remark}

	In practice, we need to choose $\eta(\cdot)$ and $g(\cdot|s_i)$ based on the available data $\{(X_i,s_i)\}_{i=1}^m$. Again, if the entire sample is used to construct $\eta(\cdot)$ and $g(\cdot|s_i)$, then the dependence among $w_i$'s and $\rho_i$'s is complicated. 
	Similar to Algorithm \ref{alg2}, we can use sample splitting to circumvent this problem, and the details are provided in Algorithm \ref{alg6}. To ensure a valid mFDR control, we require a uniformly conservative consistent estimator of $\pi(\cdot)$, whose definition is given below.
	\begin{definition}
	\label{def:unifconcon}
		An estimator $\hat{\pi}(\cdot)$ is a uniformly conservative consistent estimator of $\pi(\cdot)$, if $\sup_{i}{E}\{\hat{\pi}(s_i) - \tilde{\pi}(s_i)\}^2 \rightarrow 0$ as $m \rightarrow \infty$, where $0 \leq \tilde{\pi}(s_i) \leq \pi(s_i)$ for $i=1,\ldots,m$.
	\end{definition}
	
	The problem of constructing such uniformly conservative consistent estimator $\hat{\pi}(\cdot)$ has been discussed in the literatures; see for example, \cite{tony2019covariate}. 
	
	\begin{algorithm}[ht]
		\caption{The data-driven $\rho$-BH procedure with side information}\label{alg6}
		\begin{description}
			\item[Input:] $\{X_i\}_{i=1}^m$; $\{s_i\}_{i=1}^m$; desired mFDR level $\alpha$.
			\item[1.] Randomly split the data into two disjoint halves
			$\{X_i\}^m_{i=1}=\{X_{1,i}\}^{m_1}_{i=1}\cup\{X_{2,i}\}^{m_2}_{i=1}$, and $\{s_i\}^m_{i=1}=\{s_{1,i}\}^{m_1}_{i=1}\cup\{s_{2,i}\}^{m_2}_{i=1}$,  
			where $m_1=\lfloor m/2 \rfloor$.
			\item[2.] Use $\{X_{1,i}\}^{m_1}_{i=1}$ and $\{s_{1,i}\}^{m_1}_{i=1}$ to construct the second half alternatives estimates $\{\hat{f}_1^{\scriptscriptstyle (2)}(\cdot|s_{2,i})\}_{i=1}^{m_2}$ and a uniformly conservative consistent estimate $\hat{\pi}_2(\cdot)$.
			\item[3.] Run Algorithm \ref{alg4} with $\{X_{2,i}\}_{i=1}^{m_2}$, $\{s_{2,i}\}_{i=1}^{m_2}$, $\{\hat{f}_1^{\scriptscriptstyle (2)}(\cdot|s_{2,i})\}_{i=1}^{m_2}$, $\{\hat{\pi}_2(s_{2,i})\}_{i=1}^{m_2}$, $\{\eta(s_{2,i})\}_{i=1}^{m_2}$, $\alpha$ as inputs, where $\eta(s_{2,i}) = \hat{\pi}_2(s_{2,i})$.
			\item[4.] Switch the roles of  $\{X_{1,i}\}^{m_1}_{i=1}$,  $\{s_{1,i}\}^{m_1}_{i=1}$ and $\{X_{2,i}\}^{m_2}_{i=1}$, $\{s_{2,i}\}^{m_2}_{i=1}$. Repeat Steps 2 and 3, and combine the rejections.
			\item[Output:] The combined set of rejections.
		\end{description}
	\end{algorithm}

	The next theorem shows that Algorithm \ref{alg6} indeed controls mFDR at the target level asymptotically under conditions analogous to those assumed in Theorem \ref{thm2}. 	
	
	\begin{theorem}\label{dd:bhside}
		Assume that $\{X_i, \theta_i\}_{i=1}^m$ are independent. 
		Denote by $\{\hat{q}_{d,i}\}_{i=1}^{m_d}$, $\hat{q}_{d,(k_d)}$ and $\hat{\pi}_d$ the weighted $\rho$-values, selected thresholds and the estimated alternation proportions obtained from Algorithm \ref{alg6},  for the first and second halves of the data respectively, $d=1,2$.
		Denote by $\hat{c}_{d,i}$ the null distribution function for  $\hat{\rho}_{d,i}$. 
		Suppose $\hat{\pi}_d(s_i)>0$ and $\sup_{i}{E}\{\hat{\pi}_d(s_i) - \tilde{\pi}_d(s_i)\}^2 \rightarrow 0$ for some $\tilde\pi_d(\cdot)$ satisfying $0 \leq \tilde \pi_d(\cdot) \leq \pi(\cdot)$, and let $\tilde Q_d(t) = \frac{\sum_{i=1}^{m_d} \{1 - \tilde\pi_d(s_{d,i})\}\hat c_{d,i}(w_{d,i} t)}{{E}\{\sum_{i=1}^{m_d} \mathbb{I}(\hat q_{d,i} \leq t)\}}$ and $t_{d,L} = \sup\{t>0: \tilde Q_d(t)\leq\alpha\}$, $d=1,2$. Based on the notations from Algorithm \ref{alg6} and suppose
		\begin{enumerate}[resume* = A]
\item \label{A9} \ \ $\hat q_{d,(k_d)} \geq \nu$, $\sum_{i=1}^{m_d} P(\hat q_{d,i} \leq \nu) \geq cm$, for some constants $\nu, c>0$;
\item \label{A10} \ \ $\limsup_{t\rightarrow 0^+}\tilde Q_d(t)<\alpha$, $\liminf_{t\rightarrow \infty}\tilde Q_d(t)>\alpha$;
\item \label{A11} \ \ $\inf_{t \geq t_{d,L} + \epsilon_t}\tilde Q_d(t) \geq \alpha + \epsilon_\alpha$, $\tilde Q_d(t)$ is strictly increasing in $t\in(t_{d,L} - \epsilon_t, t_{d,L} + \epsilon_t)$, for some constants $\epsilon_\alpha, \epsilon_t > 0$.
		\end{enumerate}
 Then we have $\lim_{m\rightarrow\infty} \text{mFDR}_{\text{Algorithm } \ref{alg6}} \leq \alpha.$ 
	\end{theorem}

	\section{Numerical Experiments}\label{sec:simu}
{
In this section, we conduct several numerical experiments to compare our proposed procedures with some state-of-the-art methods. In all experiments, we study the general case where side information is available, and generate data according to the following hierarchical model:
	\begin{equation}\label{data}
			\theta_i \overset{\text{ind}}\sim \text{Ber}\{\pi(s_i)\}, \quad
			X_i | s_i, \theta_i \overset{\text{ind}}\sim (1 - \theta_i)N(0,1) + \theta_i f_1(\cdot|s_i),
	\end{equation}
	where $\theta_i \in \mathbb{R}$, $X_i \in \mathbb{R}$ and $s_i \in \mathbb{R}^l$ for $i=1,\ldots,m$. Again, we test
	$H_{0,i}:\theta_i=0$ versus $H_{1,i}:\theta_i=1$ for $i=1,\ldots,m$.
	To implement our proposed data-driven procedure with side information, i.e., Algorithm \ref{alg6}, we use the following variation of the Storey estimator to estimate $\pi(s_i)$:
	\begin{equation}\label{eq:pi}
	\hat \pi_2(s_{2,i}) = 1 - \frac{\sum_{j = 1}^{m_1} K(s_{2,i},s_{1,j})\mathbb{I}(p_{1,j} \geq \tau)}{(1-\tau)\sum_{j=1}^{m_1} K(s_{2,i},s_{1,j})}, \ \ i=1,\ldots,m_2,
	\end{equation}
	where $p_{1,j}=2\{1-\Phi(|X_{1,j}| )\}$ is the two-sided p-value with $\Phi$ being the cumulative distribution function (cdf) of standard normal variable, and $\tau$ is chosen as the p-value threshold of the BH procedure at $\alpha=0.9$; this ensures that the null cases are dominant in the set $\{j:p_{1,j}\geq \tau\}$.  Let $K(s_{2,i},s_{1,j}) = \phi_H(s_{2,i}-s_{1,j})$, where $\phi_H(\cdot)$ is the density of multivariate normal distribution with mean zero and covariance matrix $H$. We use the function 
	\texttt{Hns}
	in the R package  \texttt{ks} to chose $H$. Similar strategies for choosing  $\tau$ and $H$ are employed in \cite{cai2022laws} and \cite{ma2022napa}. 
 We construct  $\hat{f}_1^{\scriptscriptstyle (2)}(\cdot|s_{2,i})$ using a modified version of the two-step approach proposed in \cite{HART} as follows. 
 \begin{enumerate}
	\item Let $\hat \pi_1'(s_{1,i}) = 1 - \frac{\sum_{j = 1}^{m_1} K(s_{1,i},s_{1,j})\mathbb{I}(p_{1,j} \geq \tau)}{(1-\tau)\sum_{j=1}^{m_1} K(s_{1,i},s_{1,j})}$ for $i=1,\ldots,m_1$.
	\item Calculate $\tilde{f}_{1,2,j}(x_{1,j}) = \sum_{l=1}^{m_1}\frac{K(s_{1,j},s_{1,l})\phi_{h_x}(x_{1,j} - x_{1,l})}{\sum_{l=1}^{m_1}K(s_{1,j},s_{1,l})}$, \\ and the weights $\hat w_{1,j} = 1 - \min\left\{\frac{\{1-\hat \pi_1'(s_{1,j})\}f_0(x_{1,j})}{ \tilde{f}_{2,j}(x_{1,j})}, 1\right\}$ for $j=1,\ldots,m_1$.\\
	\item Obtain the non-null density estimate $\hat{f}_1^{\scriptscriptstyle (2)}(x | s_{2,i}) = \sum_{j=1}^{m_1}\frac{\hat w_{1,j}K(s_{2,i},s_{1,j}) \phi_{h_x}(x - x_{1,j})}{\sum_{j=1}^{m_1}\hat w_{1,j}K(s_{2,i},s_{1,j})}$ for $i=1,\ldots,m_2$.
 \end{enumerate}

 Here, the kernel function $K$ is the same as the one in Equation \eqref{eq:pi}, and the bandwidth $h_x$ is chosen automatically using the function \texttt{density} in the R package \texttt{stats}. 
We note that the distribution of null $\rho_i$ can be complicated, making the analytical form of $c_i(\cdot)$ intractable. Nonetheless, since we can sample from the null hypothesis $H_{0,i}$ to generate as many null copies of $\rho_i$ as needed, we are able to approximate $c_i(\cdot)$ to arbitrary accuracy. Particularly, to estimate the null densities $c_i(\cdot)$'s, i.e., the distribution functions of ${f_0(\cdot |s_{2,i})}/{\hat{f}_1^{\scriptscriptstyle (2)}(\cdot | s_{2,i})}$ under $H_{0,i}$ with $f_0$ being the density function of $N(0,1)$, $i=1,\ldots,m_2$, we independently generate 1000 samples $Y_j$'s from $f_0(\cdot|s_{2,i})$ for each $i$ and estimate $c_i(\cdot)$ through the empirical distribution of ${f_0(Y_{j}|s_{2,i})}/{\hat{f}_1^{\scriptscriptstyle (2)}(Y_{j} | s_{2,i})}$'s.
	The estimations on the first half of the data can be obtained by switching the roles of the first and the second halves. 
	
	We compare the performance of the following seven methods throughout the section:
\begin{itemize}
	\item $\rho$-BH.OR: Algorithm \ref{alg4} with $\rho_i=f_0(X_i|s_i)/f_1(X_i| s_i)$, $c_i(t)=P_{H_{0,i}}(\rho_i\leq t), \eta(\cdot)=\pi(\cdot)$.
	\item $\rho$-BH.DD: {Algorithm \ref{alg6}} with implementation details described above.
	\item LAWS: data-driven LAWS procedure \citep{cai2022laws} with p-value equals to {$2\{1-\Phi(|X_i| )\}$}. 
	\item CAMT: the CAMT procedure \citep{zhang2022covariate} with the same p-values used in LAWS.
	\item BH: the BH Procedure \citep{benjamini1995controlling} with the same p-values used in LAWS.
	\item e-BH: the e-BH Procedure \citep{wang2022false} with reciprocal of $\rho$-values used in $\rho$-BH.DD as e-values.
	\item Clfdr: {Clfdr based method \citep{HART} with Clfdr$_{d,i} = {\hat q_{d,i}}/(1+\hat q_{d,i})$, where $d=1,2$ and $\hat q_{d,i}$'s are the weighted $\rho$-values used in $\rho$-BH.DD. 
	Specifically, we calculate the threshold $k_d = \max_{1 \leq i \leq m_d}\{\sum_{j=1}^i \text{Clfdr}_{d,(j)} / i \leq \alpha\}$ and reject those with Clfdr$_{d,i} \leq \text{Clfdr}_{d, (k_d)}$.}
\end{itemize}

All simulation results are based on 100 independent replications with target level $\alpha=0.05$.
The FDR is estimated by the average of the FDP, $\sum_{i=1}^{m}\{(1-\theta_i\delta_i)/(\sum_{i=1}^{m}\delta_i\vee1)\}$, and the average power is estimated by the average proportion of the true positives that are correctly identified, $\sum_{i=1}^{m}(\theta_i\delta_i)/\sum_{i=1}^{m}\theta_i$, both over the number of repetitions.

}

\subsection{Bivariate side information}\label{sec41}
We first consider a similar setting as Setup S2 in \cite{zhang2022covariate}, where the non-null proportions and non-null distributions are closely related to a two dimensional covariate. Specifically, the parameters in Equation \eqref{data} are determined by the following equations.
\begin{equation*}
	\begin{split}
		&	s_i=(s_i^{\scriptscriptstyle (1)},s_i^{\scriptscriptstyle (2)})\overset{iid}{\sim} N(0,I_2),
		\ \  \pi(s_i) = \frac{1}{1 + e^{k_{e,i}}}, \ \ k_{e,i} = k_c + k_{d} s_i^{\scriptscriptstyle (1)},\\
		&
		 f_1(\cdot | s_i) \sim N\left(e^{k_t}{2e^{k_f s_i^{\scriptscriptstyle (2)}}}/\{1+e^{k_f s_i^{\scriptscriptstyle (2)}}\}  ,1\right), 
		 \ \ i=1,\ldots,5000,
	\end{split}
\end{equation*}
where $I_2$ is the $2\times 2$ identity matrix, $k_c$, $k_d$, $k_f$ and $k_t$ are hyper-parameters that determine the impact of $s_i$ on $\pi$ and $f_1$. In the experiments, we fix $k_c$ at 2 or 1 (denoted as ``Medium" and ``High", respectively), $(k_d, f_f)$ at $(1.5,0.4)$ or $(2.5,0.6)$ (denoted as ``Moderate" and ``Strong", respectively), and vary $k_t$ from 2 to 6. 
{\cite{zhang2022covariate} assumes it is known that $\pi(\cdot)$ and $f_1(\cdot|s_i)$ each depends on one coordinate of the covariate when implementing their procedure.}
Hence, for a fair comparison, we employ the same assumption, {substitute $s_{d,i}$ by $s_{d,i}^{\scriptscriptstyle (1)}$ for the estimations of $\hat{\pi}(\cdot)$ (as defined in \eqref{eq:pi}) and $\hat{\pi}'(\cdot)$ (as defined in Step 1 of constructing $\hat{f}_1^{\scriptscriptstyle (2)}(\cdot|s_{2,i})$), and substitute $s_{d,i}$ by $s_{d,i}^{\scriptscriptstyle (2)}$ in the rest steps of obtaining $\hat{f}_1^{\scriptscriptstyle (2)}(\cdot|s_{2,i})$, for $d=1,2$}.

	\begin{figure}[t!]
	\centering
	\includegraphics[width=0.8\textwidth]{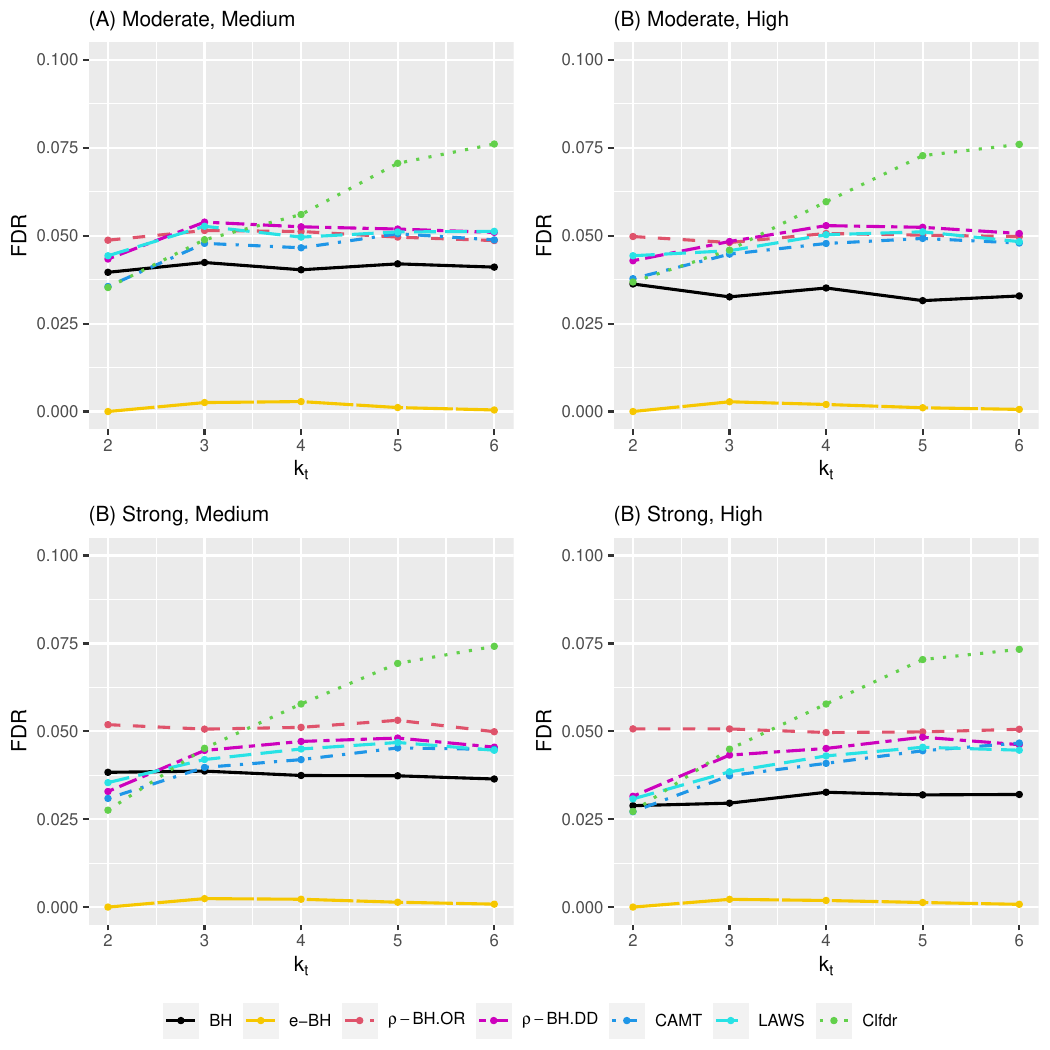}
	\caption{ The empirical FDR of BH (black solid), e-BH (yellow extra-long dash), oracle $\rho$-BH (red dashed), data-driven $\rho$-BH (purple twodash), CAMT (dark blue dotdash), LAWS (light blue longdash) and Clfdr (green dotted) for the settings described in the bivariate scenario; $\alpha=0.05$. }\label{fig:camtwpifdr}
	\end{figure}
	
	\begin{figure}[t!]
	\centering
	\includegraphics[width=0.8\textwidth]{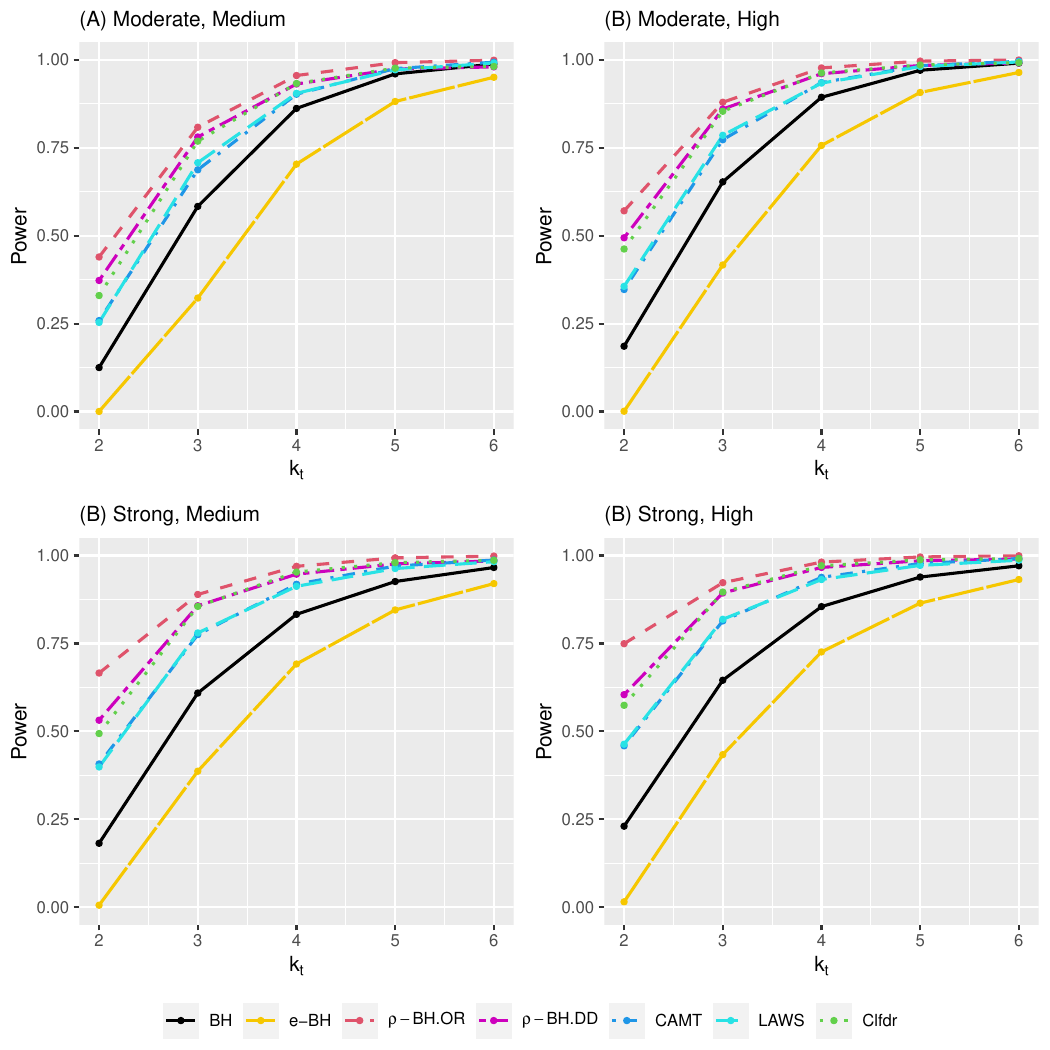}
	\caption{ The empirical power comparison, with same legend as Figure \ref{fig:camtwpifdr}.} \label{fig:camtwpipow}
	\end{figure}


It can be seen from Figure \ref{fig:camtwpifdr} that, except the Clfdr procedure, all other methods successfully control the FDR at the target level. Figure \ref{fig:camtwpipow} shows that, the empirical powers of $\rho$-BH.OR and $\rho$-BH.DD are significantly higher than all other FDR controlled methods. It is not surprising that $\rho$-BH.OR and $\rho$-BH.DD outperform LAWS and BH, because the $p$-values only rely on the null distribution, whereas the $\rho$-values mimic the likelihood ratio statistics and encode the information from the alternative distribution. 
Both $\rho$-BH.OR and $\rho$-BH.DD outperforms CAMT as well, because CAMT uses a parametric model to estimate the likelihood ratio, while $\rho$-BH.DD employs a more flexible non-parametric approach that can better capture the structural information from the alternative distribution. 
Finally, as discussed in the previous sections, the Clfdr based approaches strongly rely on the estimation accuracy of $\pi(\cdot)$ and $f_1(\cdot|\cdot)$, which can be difficult in practice. Hence as expected, we observe severe FDR distortion of Clfdr method. 
Such phenomenon reflects the advantage of the proposed $\rho$-value framework because its FDR control can still be guaranteed even if $\hat f_1(\cdot|\cdot)$ is far from the ground truth.

\subsection{Univariate side information}\label{sec42}
	Next, we consider the univariate covariate case and generate data as follows
	\begin{align*}
	\begin{split}
		&\theta_i\overset{\text{ind}}{\sim} \text{Bernoulli}\{\pi(i)\},\quad X_i|\theta_i\overset{\text{ind}}{\sim} (1-\theta_i)N(0,1)+\theta_iN(\mu,1),\quad i=1,\ldots,5000.
	\end{split}
	\end{align*}
	Two settings are considered. In Setting 1, the signals appear with elevated frequencies in the following blocks:
	$\pi(i)=0.9$ for $i\in [1001,1200]\cup[2001,2200]$; $\pi(i)=0.6$ for $i\in [3001,3200]\cup[4001,4200]$.
	For the rest of the locations we set $\pi(i)=0.01$. We vary $\mu$ from 2 to 4 to study the impact of signal strength. 
	In Setting 2, we set $\pi(i)=\pi_0$ in the above specified blocks and $\pi(i)=0.01$ elsewhere. We fix $\mu = 3$ and vary $\pi_0$ from 0.5 to 0.9 to study the influence of sparsity levels. 
	In these two cases, the side information $s_i$ can be interpreted as the signal location $i$. 
	When implementing CAMT, we use a spline basis with six equiquantile knots for $\pi(i)$ and $f_1(\cdot|i)$ to account for potential complex nonlinear effects as suggested in \cite{zhang2022covariate} and \cite{lei2018adapt}. 
	
\begin{figure}[!h]
		\centering
		\includegraphics[width=0.8\textwidth]{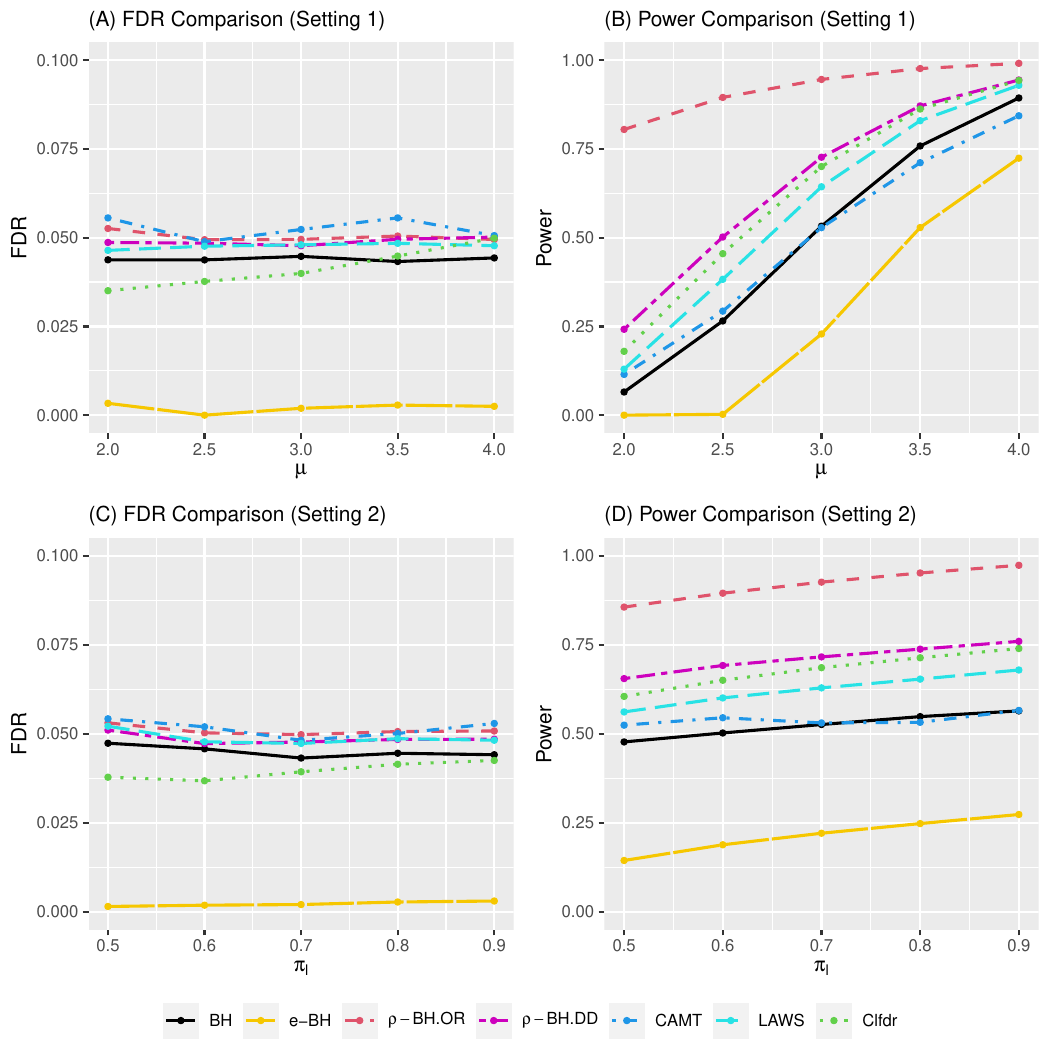}
		\caption{ The empirical FDR and power for univariate scenario, with the same legends as in Figure \ref{fig:camtwpifdr}. \label{fig:lawswpi}} 
	\end{figure}

	We compare the seven procedures as in Section \ref{sec41}, and the results of Settings 1 and 2 are summarized in the first and second rows of Figure \ref{fig:lawswpi}, respectively. 
	We can see from the first column of Figure \ref{fig:lawswpi} that, in both settings all methods control FDR appropriately at the target level. 
	From the second column of Figure \ref{fig:lawswpi}, it can be seen that both $\rho$-BH.OR and $\rho$-BH.DD outperform the other {five} methods. 
	This is due to the fact that, besides the ability in incorporating the sparsity information, the $\rho$-value statistic also adopts other structural knowledge and is henceforth more informative than the p-value based methods. 
	In addition, the nonparametric approach employed by $\rho$-BH.DD is better at capturing non-linear information than the parametric model used in CAMT, leads to a more powerful procedure. 

\section{Data Analysis}\label{sec:real}

	In this section, we compare the performance of $\rho$-BH.DD with Clfdr, CAMT, LAWS, BH, and e-BH on two real datasets.

	\subsection{MWAS data}\label{sec51}
	We first analyze a dataset from a microbiome-wide association study (MWAS) of sex effect 
\citep{mcdonald2018american}, which is available at \url{https://github.com/knightlab-analyses/american-gut-analyses}. The aim of the study is to distinguish the abundant bacteria in the gut microbiome between males and females by the sequencing of a fingerprint gene in the bacteria 16S rRNA gene. 
	This dataset is also analyzed in \cite{zhang2022covariate}. We follow their preprocessing procedure to obtain 2492 p-values from Wilcoxon rank sum test for different OTUs, and the percentages of zeros across samples for the OTUs are considered as the univariate side information. 
	Because a direct estimation of the non-null distributions of the original Wilcoxon rank sum test statistics is difficult, we construct pseudo z-values by $z_i = \Phi^{-1}(p_i) \times (2B_i-1)$, where $B_i$'s are independent $\text{Bernoulli}(0.5)$ random variables and $\Phi^{-1}$ is the inverse of standard normal cdf. 
	Then we run $\rho$-BH.DD on those pseudo z-values by employing the same estimation methods of $\pi(\cdot)$ and $f_1(\cdot|\cdot)$ as described in Section \ref{sec:simu}. 
	When implementing CAMT, we use the spline basis with six equiquantile knots as the covariates as recommended in \cite{zhang2022covariate}. 
	The results are summarized in Figure \ref{fig:realdata} (A). 
	{We can see that $\rho$-BH.DD rejects significantly more hypotheses than LAWS and BH across all FDR levels, while e-BH procedure fails to reject any hypotheses. $\rho$-BH.DD also rejects slightly more tests than Clfdr under most FDR levels, and is more stable than CAMT. 
	Because Clfdr may suffer from possible FDR inflation as shown in the simulations, we conclude that $\rho$-BH.DD enjoys the best performance on this dataset.}
	
	\subsection{ADHD data}\label{sec52}
\begin{figure}[!h]
		\centering
		\includegraphics[height=0.5\textwidth]{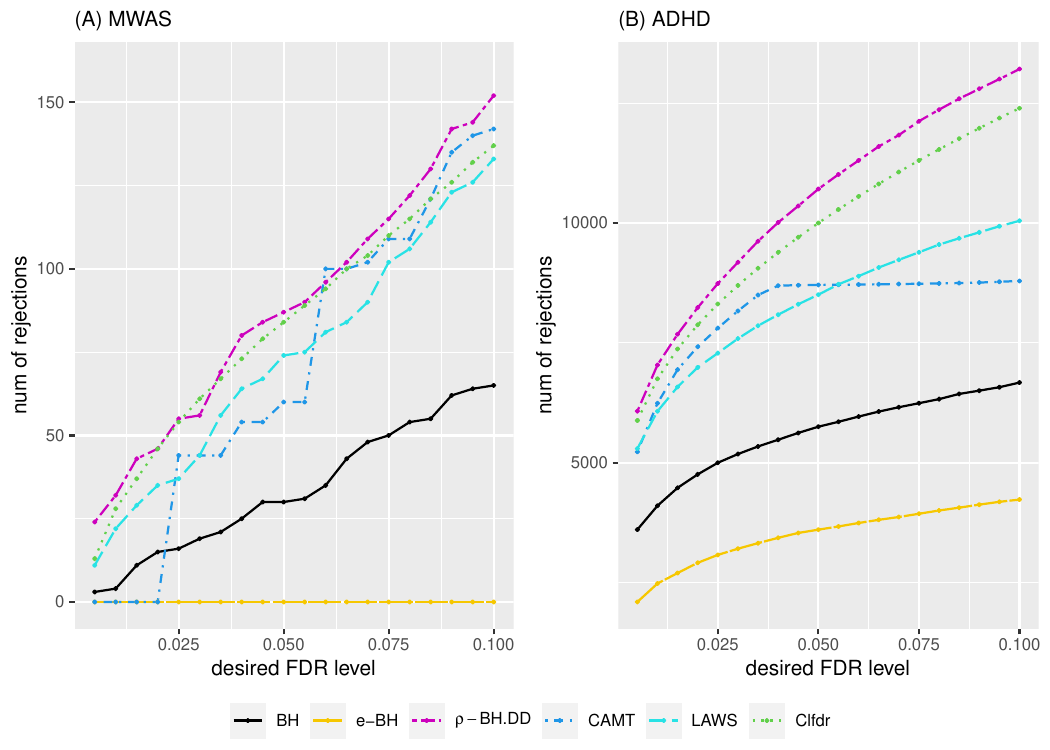}
		\caption{Total numbers of rejections for BH (black solid), e-BH (yellow extra-long dash), data-driven $\rho$-BH (purple twodash), CAMT (dark blue dotdash), LAWS (light blue longdash) and Clfdr (green dotted) at various FDR levels, respectively for MWAS (Left) and ADHD (Right) datasets. \label{fig:realdata}} 
	\end{figure}

	We next analyze a preprocessed magnetic resonance imaging (MRI) data for a study of attention deficit hyperactivity disorder (ADHD). The dataset is available at \url{http://neurobureau.projects.nitrc.org/ADHD200/Data.html}. 
	We adopt the Gaussian filter-blurred skullstripped gray matter probability images from the Athena Pipline, which are MRI images with a resolution of $197 \times 233 \times 189$.
	We pool the 776 training samples and 197 testing samples together, remove 26 samples with no ADHD index, and split the pooled data into one ADHD sample of size 585 and one normal sample of size 362. 
	Then we downsize the resolution of images to $30 \times 36 \times 30$ by taking the means of pixels within blocks, and then obtain $30 \times 36 \times 30$ two-sample t-test statistics. Similar data preprocessing strategy is also used in \cite{cai2022laws}. 
	In such dataset, the 3-dimensional coordinate indices can be employed as the side information. 
	The results of the five methods are summarized in Figure \ref{fig:realdata} (B). 
	{Again, we see that $\rho$-BH.DD rejects more hypotheses than CAMT, LAWS, Clfdr, BH and e-BH across all FDR levels.}

		\section{Discussions}\label{sec:discuss}
	This article introduces a novel multiple testing framework based on the newly proposed $\rho$-values. The strengths of this framework lie in its ability to unify existing procedures based on p-values and local false discovery rate (Lfdr) statistics, while requiring substantially weaker conditions than those typically imposed by Lfdr-based methods. Moreover, the framework naturally extends to incorporate side information through appropriate weighting schemes, under which asymptotic optimality can still be achieved.

As a concluding remark, we emphasize that the frameworks based on p-values and Lfdr statistics are not as fundamentally different as often portrayed in the literature. A central message of \cite{storey2007optimal}, \cite{sun2007oracle}, and \cite{ZAP} is that reducing z-values to p-values can result in substantial information loss, implicitly framing Lfdr and p-values as two distinct statistical paradigms. However, as we have shown in Section~\ref{sec:rhobh}, the $c(\rho)$ is a special case of a p-value but in the meantime it can yield the same ranking as the Lfdr. This suggests a more accurate interpretation of their message: Statistics that incorporate information from the alternative distribution outperform those that do not.

	To be more concrete, we show below that a Lfdr based procedure proposed in \cite{ZAP} is actually a special variation of Algorithm \ref{alg1} under Model \eqref{eq:prob}.
	Suppose $f_1$ and $f_0$ are known but $\pi$ is not. As mentioned in Section \ref{sec:dd}, a natural choice of $\hat\pi$ is the Storey estimator. Note that, the Storey estimator requires a predetermined tuning parameter $\tau$. 
		By replacing $\pi$ with $\hat{\pi}$, the threshold in the $\rho$-BH procedure becomes $\rho_{(k)}$ where
	$
		k = \max_{j} \left\{(1-\hat{\pi})c(\rho_{(j)})\leq \alpha j/m \right\} .
	$
	In a special case when we allow varying $\tau$ for different $j$ and let $\tau=c(\rho_{(j)})$, then it yields that
	{
	$
		k  =\max_{j} \left\{{\#\{i:c(\rho_i)\geq 1-c(\rho_{(j)})   \}} \leq \alpha j\right\}.
	$
	}
	Now if we add $1$ to the numerator and let
	{
	$
		k  =\max_{j} \left\{{1+\#\{i:c(\rho_i)\geq 1-c(\rho_{(j)})   \}} \leq \alpha j \right\},
	$
	}
	then the decision rule $\pmb{\delta}=\{\mathbb{I}(\rho_i\leq \rho_{(k)})\}_{i=1}^m$ is equivalent to the rule given by the ZAP procedure \citep{ZAP} that is based on Lfdr. Hence, the ZAP procedure can be viewed as a special case of the $\rho$-BH procedure under Model \eqref{eq:prob}, and can be unified into the proposed $\rho$-BH framework.

	\newpage
	\begin{appendices}

	\section{Weighted $\rho$-BH procedures}\label{ddweighted}
	In this section, we present the weighted $\rho$-BH procedures and the corresponding theoretical results. 

	Similar to incorporating prior information via a p-value weighting scheme  \citep[e.g.,][]{benjamini1997multiple,genovese2006false,dobriban2015optimal}, we can also employ such weighting strategy in the current $\rho$-value framework.  
	Let $\{w_i\}_{i=1}^m$ be a set of positive weights such that $\sum_{i=1}^{m}w_i=m$. The weighted BH procedure proposed in \cite{genovese2006false} uses $p_i/w_i$'s as the inputs of the original BH procedure. \cite{genovese2006false} proves that, {if $p_i$'s are independent and $\{w_i\}_{i=1}^m$ are independent of $\{p_i\}_{i=1}^m$ conditional on $\{\theta_i\}_{i=1}^m$}, then the weighted BH procedure controls FDR at level less than or equal to $(1-\pi)\alpha$. 
	
	Following their strategy, we can apply the adjusted weighted BH procedure (with $1-\pi$ adjustment) to $c(\rho_i)$'s and obtain the same FDR control result. However, such procedure might be suboptimal as explained in the paper.
	Alternatively, we derive a weighted $\rho$-BH procedure (without requiring $\sum_{i=1}^{m}w_i=m$) and the details are presented in Algorithm \ref{alg3}.
	
	
	\begin{algorithm}[ht]
		\caption{Weighted $\rho$-BH procedure}\label{alg3}
		\begin{description}
			\item[Input:] $\{X_i\}_{i=1}^m$; a predetermined density function $g(\cdot)$; non-null proportion $\pi$; predetermined weights $\{w_i\}_{i=1}^m$; desired FDR level $\alpha$.
			\item[1. for $i=1$ to $m$:]
			\item[] \qquad Calculate the $\rho$-values $\rho_i = f_0(X_i)/g(X_i)$.
			\item[] \qquad Compute the weighted $\rho$-values $q_i=\rho_i/w_i$.
			\item[\quad \  end for] 
			\item[2.] Sort the weighted $\rho$-values from smallest to largest  $q_{(1)} \leq \cdots \leq q_{(m)}$.
			\item[3.] Compute the null distribution function of $\rho_i$'s, and denote it by $c(\cdot)$.
			\item[4.] Let $k = \text{max}_{1 \leq j \leq m} \left\{\sum_{i=1}^{m}(1-\pi)c(q_{(j)}w_i)\leq \alpha j \right\} $.
			\item[Output:] The rejection set $\{i=1,\ldots,m: q_i\leq q_{(k)}$\}.
		\end{description}
	\end{algorithm}

	Note that $\{\rho_i/w_i\}_{i=1}^m$ in Algorithm \ref{alg3} produces a different ranking than $\{c(\rho_i)/w_i\}_{i=1}^m$, which may improve the power of the weighted p-value procedure with proper choices of $\rho$-values and weights.
	On the other hand, the non-linearity of $c(\cdot)$ imposes challenges on the theoretical guarantee for the mFDR control of Algorithm \ref{alg3} compared to that of \cite{genovese2006false}, and we derive the following result based on similar assumptions as in Theorem \ref{thm2}. 
	
	
	\begin{theorem}\label{thm3}
		Assume that $\{X_i, \theta_i\}_{i=1}^m$ are independent. Denote by $\ Q(t) = \frac{\sum_{i=1}^m (1 - \pi) c(w_i t)}{{E}\{\sum_{i=1}^m \mathbb{I}(q_i \leq t)\}}$ and $t_L = \sup\{t>0: Q(t)\leq\alpha\}.$ Based on the notations from Algorithm \ref{alg3} and suppose
		\begin{enumerate}[label=(A\arabic*), series=A, start=12, itemindent=10pt]
			\item \label{Athm5new1}  $q_{(k)} \geq \nu$ and $\sum_{i=1}^m P(q_i \leq \nu) \rightarrow \infty$ as $m\rightarrow\infty$, for some $\nu > 0$;
			\item \label{Athm5new2}  $\limsup_{t\rightarrow 0^+}Q(t)<\alpha$; $\liminf_{t\rightarrow \infty}Q(t)>\alpha$;
			\item \label{Athm5new3}  $\inf_{t \geq t_L + \epsilon_t}Q(t) \geq \alpha + \epsilon_\alpha$, $Q(t)$ is strictly increasing in $t\in(t_L - \epsilon_t, t_L + \epsilon_t)$, for some constants $\epsilon_\alpha, \epsilon_t > 0$.
		\end{enumerate}
		Then we have
		$
		\lim_{m\rightarrow\infty} \text{mFDR}_{\text{Algorithm } \ref{alg3}} \leq \alpha.
		$
	\end{theorem}
	
	
	{It is worthwhile to note that, \cite{genovese2006false} requires $\sum_{i=1}^{m}w_i=m$, which makes the weighted p-value procedure conservative. 
		In comparison, Algorithm \ref{alg3} no longer requires such condition, and it employs a tight estimate of the FDP that leads to a more powerful testing procedure. }
	
	When the oracle parameters are unknown, we can similarly construct a data-driven weighted $\rho$-BH procedure with an additional data splitting step as in Algorithm \ref{alg2}. 
We describe it in Algorithm \ref{alg5}. 


	\begin{algorithm}[ht]
		\caption{The data-driven weighted $\rho$-BH procedure}\label{alg5}
		\begin{description}
			\item[Input:] $\{X_i\}_{i=1}^m$; predetermined weights $\{w_i\}_{i=1}^m$; desired FDR level $\alpha$.
			\item[1.] Randomly split the data into two disjoint halves
			$
			\{X_i\}^m_{i=1}=\{X_{1,i}\}^{m_1}_{i=1}\cup\{X_{2,i}\}^{m_2}_{i=1}$, $\{w_i\}^m_{i=1}=\{w_{1,i}\}^{m_1}_{i=1}\cup\{w_{2,i}\}^{m_2}_{i=1}
			$, where $m_1=\lfloor m/2 \rfloor$. 
			\item[2.] Use $\{X_{1,i}\}^{m_1}_{i=1}$ to construct the second half alternative estimate $\hat{f}_1^{\scriptscriptstyle (2)}$ and a conservative consistent estimate $\hat{\pi}_2$.
			\item[3.] Run Algorithm \ref{alg3} with $\{X_{2,i}\}_{i=1}^{m_2}$, $\hat{f}_1^{\scriptscriptstyle (2)}$, $\hat{\pi}_2$, $\{w_{2,i}\}_{i=1}^{m_2}$, $\alpha$ as inputs.
			\item[4.] Switch the roles of $\{X_{1,i}\}^{m_1}_{i=1}$, $\{w_{1,i}\}^{m_1}_{i=1}$ and $\{X_{2,i}\}^{m_2}_{i=1}$, $\{w_{2,i}\}^{m_2}_{i=1}$. Repeat Steps 2 and 3, and combine the rejections.
			\item[Output:] The combined rejection set.
		\end{description}
	\end{algorithm}

The next theorem provides the theoretical guarantee for the asymptotic mFDR control of Algorithm \ref{alg5}.
\begin{theorem}\label{dd:wbh}
	Assume that $\{X_i, \theta_i\}_{i=1}^m$ are independent. 
	Denote by  $\{\hat{q}_{d,i}\}_{i=1}^{m_d}$, $\hat{q}_{d,(k_d)}$ and $\hat{\pi}_d$ the weighted $\rho$-values, selected thresholds and the estimated alternative proportions obtained from Algorithm \ref{alg5}, for the first and second halves of the data respectively, $d=1,2$.
	Denote by $\hat{c}_{d}$ the null distribution function for $\hat{\rho}_{d,i}$.
	Suppose $\hat{\pi}_d>0$ and $|\hat{\pi}_d - \tilde{\pi}_d|\overset{P}{\rightarrow} 0$ for some $\tilde{\pi}_d$ satisfying $0\leq\tilde{\pi}_d\leq\pi$, and let $\tilde Q_d(t) = \frac{\sum_{i=1}^{m_d} (1 - \tilde\pi_d)\hat c_{d}(w_{d,i} t)}{ E \{\sum_{i=1}^{m_d} \mathbb{I}(\hat q_{d,i} \leq t)\}}$ and $t_{d,L} = \sup\{t>0: \tilde Q_d(t)\leq\alpha\}$, $d=1,2$. Based on the notations from Algorithm \ref{alg5} and suppose 
	\begin{enumerate}[label=(A\arabic*), series=A, start=15, itemindent=10pt]
		\item \label{A12}  $\hat q_{d,(k_d)} \geq \nu$, $\sum_{i=1}^{m_d}  P (\hat q_{d,i} \leq \nu) \geq cm$, for some constants $\nu, c>0$;
		\item \label{A13}  $\limsup_{t\rightarrow 0^+}\tilde Q_d(t)<\alpha$, $\liminf_{t\rightarrow \infty}\tilde Q_d(t)>\alpha$;
		\item \label{A14}  $\inf_{t \geq t_{d,L} + \epsilon_t}\tilde Q_d(t) \geq \alpha + \epsilon_\alpha$, $\tilde Q_d(t)$ is strictly increasing in $t\in(t_{d,L} - \epsilon_t, t_{d,L} + \epsilon_t)$, for some constants $\epsilon_\alpha, \epsilon_t > 0$.
	\end{enumerate} 
	Then we have $\lim_{m\rightarrow\infty} \text{mFDR}_{\text{Algorithm } \ref{alg5}} \leq \alpha.$
\end{theorem}

\section{Proofs of Main Theorems and Propositions}\label{sec:thmpf}

Note that Theorems \ref{thm1} and \ref{thm:by} follow directly from the proof of the original BH procedure as discussed in the main text. Theorem \ref{thm:opt1} is a special case of Theorem \ref{thm:opt2}. Theorems \ref{thm2} and \ref{dd:wbh} are special cases of Theorem \ref{dd:bhside} with slight modifications. {Theorem \ref{thm3} is a special case of Theorem \ref{thm4}.}
Hence, we focus on the proofs of Proposition \ref{prop1}, Theorem \ref{thm4}, Theorem \ref{thm:opt2} and Theorem \ref{dd:bhside} in this section.

To simplify notation, we define a  procedure equivalent to Algorithm \ref{alg4}. This equivalence is stated in Lemma \ref{lemma:eqv}, whose proof will be given later in \ref{lemmas}.

\begin{algorithm}
	\caption{A procedure equivalent to Algorithm \ref{alg4}}\label{alg:eqv}
	\begin{description}
		\item[Input:] $\{X_i\}_{i=1}^m$; $\{s_i\}_{i=1}^m$; predetermined density functions $\{g(\cdot|s_i)\}_{i=1}^m$; non-null proportions $\{\pi(s_i)\}_{i=1}^m$; predetermined $\{\eta(s_i)\}_{i=1}^m$; desired FDR level $\alpha$.
		\item[1. for $i=1$ to $m$:]
		\item[] \qquad Calculate the $\rho$-values $\rho_i = f_0(X_i|s_i)/g(X_i|s_i)$.
		\item[] \qquad Compute the null distribution functions of each $\rho_i$, and denote them by $\{c_i(\cdot)\}_{i=1}^m$.
		\item[] \qquad Let $w_i=\frac{\eta(s_i)}{1-\eta(s_i)}$, and compute the weighted $\rho$-values $q_i=\rho_i/w_i$, $i=1,\ldots,m$.
		\item[\qquad\ end for]
		\item[2.]  Let $t^* = \text{max}_{t \geq 0} \left\{\frac{\sum_{i=1}^{m} \{1-\pi(s_i)\}c_i(w_i t)}{\{\sum_{i=1}^{m} \mathbb{I}(q_i \leq t)\} \lor 1}\leq \alpha \right\} $.
		\item[Output:] The set of rejections $\{i=1,\ldots,m: q_i \leq t^*\}$.
	\end{description}
\end{algorithm}

\begin{lemma}
	\label{lemma:eqv}
		Algorithm \ref{alg4} and Algorithm \ref{alg:eqv} are equivalent in the sense that
	they reject the same set of hypotheses.
\end{lemma}

\subsection{Proof of Proposition \ref{prop1}}
\begin{proof}
	Denote by $\hat{ P }\{c(\rho_i)>\tau\}:=\dfrac{\sum_{i=1}^{m}\mathbb{I}\{c(\rho_i)>\tau\}}{m}$. Since $\sum_{i=1}^{m}\mathbb{I}\{c(\rho_i)>\tau\}$ follows $\text{Binomial}(m,p)$ where $p= P \{c(\rho_i)>\tau\}$, we have that 
	$$
	\hat{ P }\{c(\rho_i)>\tau\}\overset{P}{\rightarrow} p.
	$$
	Let $p_0= P \{c(\rho_i)>\tau|H_{0,i}\}$ and $p_1= P \{c(\rho_i)>\tau|H_{1,i}\}$, then $p=(1-\pi)p_0+\pi p_1$. Since $c(\rho_i)\sim \text{Unif}(0,1)$ under $H_{0,i}$, it follows that $p=(1-\pi)(1-\tau)+\pi p_1$. Hence, $1-p/(1-\tau)<\pi$, and the proposition follows.
\end{proof}
		
	\subsection{Proof of Theorem \ref{thm4}}
				
	\begin{proof}
		By Lemma \ref{lemma:eqv}, we only need to prove the mFDR control for Algorithm \ref{alg:eqv}. Assumption \ref{A5} ensures that $Q(t)$ is well defined when $t \geq \nu$. Note that, by Assumption \ref{A5} and standard Chernoff bound for independent Bernoulli random variables, we have uniformly for $t \geq \nu$ and any $\epsilon > 0$
	\begin{equation*}
		\begin{split}
			& P  \left( \left| \frac{\sum_{i=1}^m \mathbb{I}(q_i \leq t)}{ E \{\sum_{i=1}^m \mathbb{I}(q_i \leq t)\}} -1\right| \geq \epsilon \right) \\
			\leq & 2e^{-\epsilon^2\sum_{i=1}^{m}  P (q_i \leq t)/3} \\
			\leq & 2e^{-\epsilon^2\sum_{i=1}^{m}  P (q_i \leq \nu)/3} 
			\rightarrow  0,
		\end{split}
	\end{equation*}
	which implies
	\begin{equation}
	\label{eq:thm6:1}
		\sup_{t\geq\nu} |Q(t) - \widehat{\text{FDP}}(t)| \overset{P}{\rightarrow} 0
	\end{equation}
 as $m\rightarrow\infty$, where $\widehat{\text{FDP}}(t) = \frac{\sum_{i=1}^m \{1 - \pi(s_i)\} c_i(w_i t)}{\{\sum_{i=1}^m \mathbb{I}(q_i \leq t)\} \lor 1}.$
	
	Assumption \ref{A6} implies $t_L < \infty$. Moreover, combining Equation \eqref{eq:thm6:1} with Assumption \ref{A7}, we have $\widehat{\text{FDP}}(t) > \alpha$ for any $t \geq t_L+\epsilon_t$ with probability going to $1$. Thus, we only have to consider $t < t_L+\epsilon_t$. Specifically, we consider $t\in(t_L-\epsilon_t, t_L+\epsilon_t)$. As $Q(t)$ is strictly increasing within this range by Assumption \ref{A7}, we have
	\begin{equation*}
		t^* = Q^{-1}\{Q(t^*)\} \overset{P}{\rightarrow} Q^{-1}\{\widehat{\text{FDP}}(t^*)\} = Q^{-1}(\alpha) = t_L.
	\end{equation*}
		
	Therefore, we have
		\begin{equation}
			\begin{split}
				\text{mFDR}_{\text{Algorithm }  \ref{alg:eqv}} & = \frac{\sum_{i=1}^m  P (q_i \leq t^*, \theta_i = 0)}{ E \{\sum_{i=1}^m \mathbb{I}(q_i \leq t^*)\}} \\
				&= \frac{\sum_{i=1}^m  P (q_i \leq t_L, \theta_i = 0)}{ E \{\sum_{i=1}^m \mathbb{I}(q_i \leq t_L)\}} + o(1)\\
				& = Q(t_L) + o(1) \leq \alpha + o(1).
			\end{split}
		\end{equation}
	
		
	\end{proof}

	\subsection{Proof of Theorem \ref{thm:opt2}}
    
        We first state a useful lemma whose proof will be given later in \ref{lemmas}.
        \begin{lemma}
            \label{lemma2}
            Let $g(\cdot|s_i) \equiv f_1(\cdot|s_i)$, $i=1,\ldots,m$, $\eta(\cdot) \equiv \pi(\cdot)$. For any $t > 0$, let
            \begin{equation*}
                Q(t) = \frac{\sum_{i=1}^m \{1 - \pi(s_i)\} c_i(w_i t)}{ E \{\sum_{i=1}^m \mathbb{I}(q_i \leq t)\}}, \quad t_L = \text{sup} \{ t \in (0, \infty): Q(t) \leq \alpha \}.
            \end{equation*}
            
            Suppose Assumption \ref{A8} holds. Then we have
    
            \begin{enumerate}
                \item $Q(t) < \frac{t}{1+t}$;
                \item $Q(t)$ is strictly increasing;
                \item $\lim_{m\rightarrow\infty} (\text{ETP}_{\bm \delta^L} -  \text{ETP}_{\bm \delta'}) \geq 0$, for any testing rule $\bm \delta'$ based on $\{X_i\}_{i=1}^m$ and $\{s_i\}_{i=1}^m$ such that $\lim_{m\rightarrow\infty}\text{mFDR}_{\bm \delta'} \leq \alpha$, where $\bm \delta^L = \{\mathbb{I}(q_i \leq t_L)\}_{i=1}^m$.
            \end{enumerate}		
        \end{lemma}
    
        Next we prove Theorem \ref{thm:opt2}.
    
        \begin{proof}
            By Lemma \ref{lemma:eqv}, Algorithm \ref{alg4} is equivalent to reject all hypotheses that satisfying $q_i \leq t^*$, where $t^*$ is the threshold defined in Algorithm \ref{alg:eqv}. To simplify notations, let $\nu = \frac{\alpha}{1-\alpha}$ and we next show that $t^* \geq \nu$ in probability.
            
            By the standard Chernoff bound for independent Bernoulli random variables, we have
            \begin{equation*}
                \begin{split}
                     P  \left( \left| \frac{\sum_{i=1}^{m} \mathbb{I}(q_i \leq \nu)}{\sum_{i=1}^{m}  P (q_i \leq \nu)} - 1\right| \geq \epsilon \right) \leq 2e^{-\epsilon^2\sum_{i=1}^{m}  P (q_i \leq \nu)/3}
                \end{split}
            \end{equation*}
            for all $0 < \epsilon < 1$. By Assumption \ref{A8}, the above implies
            \begin{equation}
            \label{eq:thm7:t>nu1}
                \left| \frac{\sum_{i=1}^{m} \mathbb{I}(q_i \leq \nu)}{\sum_{i=1}^{m}  P (q_i \leq \nu)} - 1\right| = o_P(1).
            \end{equation}
            
            Combining Equation \eqref{eq:thm7:t>nu1} and the first part of Lemma \ref{lemma2}, we have
            \begin{equation}
            \label{eq:thm7:t>nu3}
                \begin{split}
                    \frac{\sum_{i=1}^{m} \{1-\pi(s_i)\}c_i(w_i \nu)}{\{\sum_{i=1}^{m} \mathbb{I}(q_i \leq \nu)\} \lor 1}
                    =& \frac{\sum_{i=1}^{m} \{1-\pi(s_i)\}c_i(w_i \nu)}{\sum_{i=1}^{m}  P (q_i \leq \nu)} + o_P(1)\\
                    =& Q(\nu) + o_P(1)
                    <  \frac{\nu}{1+\nu} + o_P(1)\\
                    =&  \alpha + o_P(1),
                \end{split}
            \end{equation}
            which implies $ P (t^* \geq \nu) \rightarrow 1$. Therefore, we will only focus the event $\{t^* \geq \nu\}$ in the following proof.
            
            For any $t > 0$, we let
            \begin{equation*}
            \begin{split}
                &Q(t) = \frac{\sum_{i=1}^m \{1 - \pi(s_i)\} c_i(w_i t)}{ E \{\sum_{i=1}^m \mathbb{I}(q_i \leq t)\}},\\
                &\widehat{\text{FDP}}(t) = \frac{\sum_{i=1}^m \{1 - \pi(s_i)\} c_i(w_i t)}{\{\sum_{i=1}^m \mathbb{I}(q_i \leq t)\} \lor 1},
            \end{split}
            \end{equation*}
            and
            \begin{equation*}
            \begin{split}
                &t_L = \text{sup} \{ t \in (0, \infty): Q(t) \leq \alpha \}, \\ 
                &t^* = \text{sup} \{ t \in (0, \infty): \widehat{\text{FDP}}(t) \leq \alpha \}.
            \end{split}
            \end{equation*}
    
            Following the proof of the third part of Lemma \ref{lemma2}, we consider two cases: $\lim\limits_{m\rightarrow \infty}\frac{\pi(s_i)}{m} \leq 1-\alpha$ and $\lim\limits_{m\rightarrow\infty}\frac{\pi(s_i)}{m} > 1-\alpha$. 
            
            The first case is trivial by noting that mFDR can be controlled even if we reject all null hypotheses.
            For the second case, we need to show that $t^* {\overset{P}{\rightarrow}} t_L$. Similar to the proof of Equation \eqref{eq:thm7:t>nu1}, we have uniformly for $t\geq\nu$ and any $\epsilon > 0$,
            \begin{equation*}
                \begin{split}
                    & P \left(\left| \frac{\sum_{i=1}^m \mathbb{I}(q_i \leq t)\}}{ E \{\sum_{i=1}^m \mathbb{I}(q_i \leq t)\}} - 1 \right| \geq \epsilon \right) \\
                    \leq & 2e^{-\epsilon^2\sum_{i=1}^{m}  P (q_i \leq t)/3} \\
                    \leq & 2e^{-\epsilon^2\sum_{i=1}^{m}  P (q_i \leq \nu)/3} 
                    \rightarrow  0,
                \end{split}
            \end{equation*}
            which implies $|\widehat{\text{FDP}}(t) - Q(t)| \overset{P}{\rightarrow} 0$ uniformly in $t \geq \nu$. Thus, $\widehat{\text{FDP}}(t^*) \overset{P}{\rightarrow} Q(t^*)$. Moreover, by Lemma \ref{lemma2}, we know that $Q(t)$ is continuous and strictly increasing. Therefore, we can define the inverse function $Q^{-1}(\cdot)$ of $Q(\cdot)$. Thus, by the continuous mapping theorem, we have 
            \begin{equation*}
                t^* = Q^{-1}\{Q(t^*)\} \overset{P}{\rightarrow} Q^{-1}\{\widehat{\text{FDP}}(t^*)\} = Q^{-1}(\alpha) = t_L.
            \end{equation*}
                {By the third part of Lemma \ref{lemma2}, we have $\lim\limits_{m\rightarrow\infty}(\text{ETP}_{\bm \delta^L} - \text{ETP}_{\bm \delta}) \geq 0$ and therefore, 
            \begin{equation*}
                \begin{split}
                    \lim_{m\rightarrow\infty}\frac{\text{ETP}_{\bm \delta_{\rho}}}{\text{ETP}_{\bm \delta}} 
                    &=\lim_{m\rightarrow\infty}\frac{\text{ETP}_{\bm \delta_{\rho}}}{\text{ETP}_{\bm \delta_L}} \frac{\text{ETP}_{\bm \delta_L}}{\text{ETP}_{\bm \delta}}
                    \geq \lim_{m\rightarrow\infty}\frac{\text{ETP}_{\bm \delta_{\rho}}}{\text{ETP}_{\bm \delta_L}}\\
                    &=\lim_{m\rightarrow\infty}\frac{ E \{\sum_{i=1}^m \theta_i\mathbb{I}(\rho_i \leq t^*)\}}{ E \{\sum_{i=1}^m \theta_i\mathbb{I}(\rho_i \leq t_L)\}}\\
                    &\geq1 + \lim_{m\rightarrow\infty}\frac{\sum_{i=1}^m \{1-\pi(s_i)\} o(1)}{\sum_{i=1}^m \{1-\pi(s_i)\}c_i(w_i t_L)} \geq 1.\\
                \end{split}
            \end{equation*}
            }
            \end{proof}

        \subsection{Proof of Theorem \ref{dd:bhside}}
        
        We first introduce Lemma \ref{lemma3}, whose proof will be given later in \ref{lemmas}.
        \begin{lemma}
            \label{lemma3}
            Denote Steps 2 to 3 of Algorithm \ref{alg6} as `Half-procedure' and we inherit all other notations from Theorem \ref{dd:bhside}. Suppose Assumptions \ref{A9}-\ref{A11} hold for $d=2$. Then we have
            \begin{equation*}
                \lim_{m\rightarrow\infty} \text{mFDR}_{\text{Half-procedure}} \leq \alpha.
            \end{equation*}
        \end{lemma}
    Next we prove Theorem \ref{dd:bhside}.
        \begin{proof}		
        Without loss of generality, we assume $\{X_{1,i}\}^{m_1}_{i=1} = \{X_i\}^{m_1}_{i=1}$ and $\{X_{2,i}\}^{m_2}_{i=1} = \{X_i\}^{m}_{i=m_1+1}$. By Lemma \ref{lemma:eqv} and Lemma \ref{lemma3}, we have that
            \begin{equation}
                \label{eq:thm8:1}
                \begin{split}
                    &\frac{ E  \{\sum_{i=1}^{m_1} (1-\theta_i)\delta_i\}}{ E  \{\sum_{i=1}^{m_1} \delta_i\}} \leq \alpha + o(1),\\
                    &\frac{ E  \{\sum_{i=m_1+1}^{m} (1-\theta_i)\delta_i\}}{ E  \{\sum_{i=m_1+1}^{m} \delta_i\}} \leq \alpha + o(1).
                \end{split}
            \end{equation}
            On the other hand, we can decompose $\text{mFDR}_{\bm \delta}$ as
            \begin{equation}
                \label{eq:thm8:2}
                \begin{split}
                    \text{mFDR}_{\bm \delta} =& \frac{ E  \{\sum_{i=1}^{m} (1-\theta_i)\delta_i\}}{ E  \{\sum_{i=1}^{m} \delta_i\}} \\
                    =& \frac{E \{\sum_{i=1}^{m_1} (1-\theta_i)\delta_i\}}{E \{\sum_{i=1}^{m} \delta_i\}} + \frac{E \{\sum_{i=m_1+1}^{m} (1-\theta_i)\delta_i\}}{E \{\sum_{i=1}^{m} \delta_i\}} \\
                    =& \frac{E \{\sum_{i=1}^{m_1} (1-\theta_i)\delta_i\}}{E \{\sum_{i=1}^{m_1} \delta_i\}} \frac{E \{\sum_{i=1}^{m_1} \delta_i\}}{E \{\sum_{i=1}^{m} \delta_i\}} +\\ &\frac{E \{\sum_{i=m_1+1}^{m} (1-\theta_i)\delta_i\}}{E \{\sum_{i=m_1+1}^{m} \delta_i\}} \frac{E \{\sum_{i=m_1+1}^{m} \delta_i\}}{E \{\sum_{i=1}^{m} \delta_i\}}.
                \end{split}
            \end{equation}
            
            Therefore, by Equations \eqref{eq:thm8:1} and \eqref{eq:thm8:2}, we conclude that
            \begin{equation*}
                \begin{split}
                    \lim_{m\rightarrow\infty} \text{mFDR}_{\bm \delta} &\leq \alpha \left\{\frac{E (\sum_{i=1}^{m_1} \delta_i)}{E (\sum_{i=1}^{m} \delta_i)} + \frac{E (\sum_{i=m_1+1}^{m} \delta_i)}{E (\sum_{i=1}^{m} \delta_i)} \right\} 
                    = \alpha.
                \end{split}
            \end{equation*}
        \end{proof}

        \section{Proofs of Lemmas}\label{lemmas}	
        
        \subsection{Proof of Lemma \ref{lemma:eqv}}
        
        \begin{proof}
            It is easy to see that $t^* \geq q_{(k)}$ as
            \begin{equation*}
                \frac{\sum_{i=1}^{m} \{1-\pi(s_i)\}c_i(w_i q_{(k)})}{\sum_{i=1}^{m} \mathbb{I}(q_i \leq q_{(k)})}\leq \alpha.
            \end{equation*}
            
            Now it suffices to show that, for any $t \geq q_{(k+1)}$, we have
            \begin{equation}
                \label{eq:lemma1:p1:1}
                \frac{\sum_{i=1}^{m} \{1-\pi(s_i)\}c_i(w_i t)}{\sum_{i=1}^{m} \mathbb{I}(q_i \leq t)}> \alpha.
            \end{equation}
            
            By the definition of $k$, for any $l \geq k+1$, we have 
            \begin{equation*}
                \frac{\sum_{i=1}^{m} \{1-\pi(s_i)\}c_i(w_i q_{(l)})}{\sum_{i=1}^{m} \mathbb{I}(q_i \leq q_{(l)})} > \alpha.
            \end{equation*}
            
            Then for any $l \geq k+1$, for any $t \in [q_{(l)},q_{(l+1)})$ where $q_{(m+1)} = \infty$, we have
            \begin{equation*}
                \begin{split}
                    \frac{\sum_{i=1}^{m} \{1-\pi(s_i)\}c_i(w_i t)}{\sum_{i=1}^{m} \mathbb{I}(q_i \leq t)} 
                    = & \frac{\sum_{i=1}^{m} \{1-\pi(s_i)\}c_i(w_i t)}{l} \\
                    \geq & \frac{\sum_{i=1}^{m} \{1-\pi(s_i)\}c_i(w_i q_{(l)})}{l} \\
                    = & \frac{\sum_{i=1}^{m} \{1-\pi(s_i)\}c_i(w_i q_{(l)})}{\sum_{i=1}^{m} \mathbb{I}(q_i \leq q_{(l)})} 
                    >  \alpha.
                \end{split}
            \end{equation*}
            
            This proves Equation \eqref{eq:lemma1:p1:1} and concludes the proof.
            
        \end{proof}

        \subsection{Proof of Lemma \ref{lemma2}}
        \begin{proof}		
            First of all, by Assumption \ref{A8}, we have that $Q(t)$ is well defined for $t \geq \nu$. For any $t$ such that $ E \{\sum_{i=1}^m\mathbb{I}(q_i\leq t)\}=0$, we set $Q(t)=0$ for simplicity and it will not affect the results. We can rewrite $Q(t)$ as
            \begin{equation}
                \label{eq:lemma3:1}
                \begin{split}
                    Q(t) &= \frac{ E \{\sum_{i=1}^m(1-\theta_i)\delta_i\}}{ E (\sum_{i=1}^m\delta_i)} 
                    = \frac{ E [\sum_{i=1}^m  E \{ (1-\theta_i)\delta_i | X_i\}]}{ E (\sum_{i=1}^m\delta_i)} \\
                    &= \frac{ E [\sum_{i=1}^m \delta_i  E \{ (1-\theta_i) | X_i\}]}{ E (\sum_{i=1}^m\delta_i)} 
                    = \frac{ E \left\{\sum_{i=1}^m \mathbb{I}( q_i \leq t) \frac{q_i}{1+q_i}\right\}}{ E \{\sum_{i=1}^m\mathbb{I}( q_i \leq t)\}}.
                \end{split}
            \end{equation}
                For the first part of this lemma, note that
            \begin{equation*}
                \begin{split}
                    & E \left\{\sum_{i=1}^m \mathbb{I}(q_i \leq t) \frac{q_i}{1+q_i}\right\}-\frac{t}{1+t} E \left\{\sum_{i=1}^m\mathbb{I}(q_i \leq t)\right\} \\
                    &= E \left\{\sum_{i=1}^m \mathbb{I}(q_i \leq t) \left(\frac{q_i}{1+q_i} - \frac{t}{1+t}\right)\right\} \\
                    &=  E \left\{\sum_{i=1}^m \mathbb{I}(q_i \leq t) \frac{q_i-t}{(1+q_i)(1+t)}\right\} 
                    \leq 0.
                \end{split}
            \end{equation*}
            The equality holds if and only if $ P (q_i < t | q_i \leq t) = 0$. Therefore, by Equation \eqref{eq:lemma3:1}, we have
            \begin{equation}
            \label{eq:lemma2:tl>nu1}
                Q(t) = \frac{ E \left\{\sum_{i=1}^m \mathbb{I}(q_i \leq t) \frac{q_i}{1+q_i}\right\}}{ E \{\sum_{i=1}^m\mathbb{I}(q_i \leq t)\}} < \frac{t}{1+t}.
            \end{equation}
            Denote by $\nu = \frac{\alpha}{1-\alpha}$. By Equation \eqref{eq:lemma2:tl>nu1}, we immediately have $t_L \geq \nu$. Therefore, we only consider $t \geq \nu$ in the following proof.
            
            For the second part, let $\nu \leq t_1 < t_2 < \infty$, $Q(t_1) = \alpha_1$ and $Q(t_2) = \alpha_2$. From the first part, we learn that $\alpha_1 < \frac{t_1}{1+t_1}$. Therefore, 
            \begin{equation*}
                \begin{split}
                    Q(t_2) =& \frac{ E \left\{\sum_{i=1}^m \mathbb{I}(q_i \leq t_2) \frac{q_i}{1+q_i}\right\}}{ E \left\{\sum_{i=1}^m\mathbb{I}(q_i \leq t_2)\right\}} \\
                    =& \frac{ E \left\{\sum_{i=1}^m \mathbb{I}(q_i \leq t_1) \frac{q_i}{1+q_i}\right\}}{ E \{\sum_{i=1}^m\mathbb{I}(q_i \leq t_2)\}} + \frac{ E \left\{\sum_{i=1}^m \mathbb{I}(t_1 < q_i \leq t_2) \frac{q_i}{1+q_i}\right\}}{ E \{\sum_{i=1}^m\mathbb{I}(q_i \leq t_2)\}} \\
                    =& \frac{ E \left\{\sum_{i=1}^m \mathbb{I}(q_i \leq t_1) \frac{q_i}{1+q_i}\right\}}{ E \{\sum_{i=1}^m\mathbb{I}(q_i \leq t_1)\}} \frac{ E \{\sum_{i=1}^m\mathbb{I}(q_i \leq t_1)\}}{ E \{\sum_{i=1}^m\mathbb{I}(q_i \leq t_2)\}}+ \\
                    &\frac{ E \left\{\sum_{i=1}^m \mathbb{I}(t_1 < q_i \leq t_2) \frac{q_i}{1+q_i}\right\}}{ E \{\sum_{i=1}^m\mathbb{I}(q_i \leq t_2)\}} \\
                    =& \alpha_1 \frac{ E \{\sum_{i=1}^m\mathbb{I}(q_i \leq t_1)\}}{ E \{\sum_{i=1}^m\mathbb{I}(q_i \leq t_2)\}} + \frac{ E \left\{\sum_{i=1}^m \mathbb{I}(t_1 < q_i \leq t_2) \frac{q_i}{1+q_i}\right\}}{ E \{\sum_{i=1}^m\mathbb{I}(q_i \leq t_2)\}}\\
                    \geq& \alpha_1 \frac{ E \{\sum_{i=1}^m\mathbb{I}(q_i \leq t_1)\}}{ E \{\sum_{i=1}^m\mathbb{I}(q_i \leq t_2)\}} + \frac{t_1}{1+t_1}\frac{ E \{\sum_{i=1}^m \mathbb{I}(t_1 < q_i \leq t_2)\}}{ E \{\sum_{i=1}^m\mathbb{I}(q_i \leq t_2)\}} \\
                    >& \alpha_1 \frac{ E \{\sum_{i=1}^m\mathbb{I}(q_i \leq t_1)\}}{ E \{\sum_{i=1}^m\mathbb{I}(q_i \leq t_2)\}} + \alpha_1 \frac{ E \{\sum_{i=1}^m \mathbb{I}(t_1 < q_i \leq t_2)\}}{ E \{\sum_{i=1}^m\mathbb{I}(q_i \leq t_2)\}} \\
                    =& \alpha_1 = Q(t_1).
                \end{split}
            \end{equation*}
            
            For the third part, note that $Q(t)$ here is continuous and increasing when $m \rightarrow \infty$. We consider two cases: $\lim\limits_{m\rightarrow\infty}\frac{\pi(s_i)}{m} \leq 1-\alpha$ and $\lim\limits_{m\rightarrow\infty}\frac{\pi(s_i)}{m} > 1-\alpha$. 
            
            The first case is trivial since it implies $\lim\limits_{t\rightarrow\infty} Q(t) \leq \alpha$ and $t_L = \infty$. The procedure rejects all hypotheses and is obviously most powerful.
            For the second case, we have $\lim_{t\rightarrow\infty} Q(t) = \frac{\sum_{i=1}^m \{1-\pi(s_i)\}}{m} > \alpha$. Combining this with the fact that $Q(\nu) < \alpha$, we can always find a unique $t_L$ such that $Q(t_L) = \alpha$. Note that, by 
            \begin{equation*}
            \begin{split}
                & \lim_{m\rightarrow\infty}\text{mFDR}_{\bm \delta^L} = \lim_{m\rightarrow\infty}\frac{ E \left\{\sum_{i=1}^m \mathbb{I}( q_i \leq t_L) \frac{q_i}{1+q_i}\right\}}{ E \{\sum_{i=1}^m\mathbb{I}( q_i \leq t_L)\}} = \alpha,\\
                \text{and} \ &\lim_{m\rightarrow\infty}\text{mFDR}_{\bm \delta'} = \lim_{m\rightarrow\infty}\frac{ E \left\{\sum_{i=1}^m \delta_i'\frac{q_i}{1+q_i}\right\}}{\{ E \sum_{i=1}^m \delta_i'\}} \leq \alpha,
            \end{split}
            \end{equation*}
            we have
            \begin{align*}
            \begin{split}
                &\lim_{m\rightarrow\infty} E \left\{\sum_{i=1}^m \delta_i^L\left(\frac{q_i}{1+q_i} - \alpha\right)\right\}=0 \\  \text{and}\  &\lim_{m\rightarrow\infty} E \left\{\sum_{i=1}^m \delta_i'\left(\frac{q_i}{1+q_i} - \alpha\right)\right\}\leq 0,
            \end{split}
            \end{align*}
            which implies 
            \begin{equation}
                \label{eq:lemma3:2}
                \lim_{m\rightarrow\infty}  E \left\{\sum_{i=1}^m (\delta_i^L - \delta_i')\left(\frac{q_i}{1+q_i} - \alpha\right)\right\} \geq 0.
            \end{equation}
            Note that, by the law of total expectation as in Equation \eqref{eq:lemma3:1}, we have
            \begin{align}
                \label{eq:lemma3:3}
                \begin{split}
                &\lim_{m\rightarrow\infty}\{ E (\sum_{i=1}^m \delta_i^L \theta_i) -  E (\sum_{i=1}^m \delta_i' \theta_i)\} \geq 0 \\ \Leftrightarrow \  &\lim_{m\rightarrow\infty} E \left\{\sum_{i=1}^m (\delta_i^L - \delta_i')\frac{1}{1+q_i}\right\} \geq 0.
            \end{split}
            \end{align}
            Hence, it suffices to show $\lim_{m\rightarrow\infty} E \left\{\sum_{i=1}^m (\delta_i^L - \delta_i')\frac{1}{1+q_i}\right\} \geq 0.$ By Equation \eqref{eq:lemma3:2}, it suffices to show that there exists  some $\lambda \geq 0$ such that $(\delta_i^L - \delta_i')\frac{1}{1+q_i} \geq \lambda (\delta_i^L - \delta_i')\left(\frac{q_i}{1+q_i} - \alpha\right)$ for every $i$, i.e.,
            \begin{equation}
                \label{eq:lemma3:4}
                (\delta_i^L - \delta_i')\left\{\frac{1}{1+q_i} - \lambda\left(\frac{q_i}{1+q_i} - \alpha\right)\right\} \geq 0.
            \end{equation} 
            
            By the first part of this lemma, we have $\alpha = Q(t_L) < \frac{t_L}{1+t_L}$ and thus $\frac{1}{t_L - \alpha(1+t_L)} > 0$. 
            Let $\lambda = \frac{1}{t_L - \alpha(1+t_L)}$, then for each $i$: 		
            \begin{enumerate}
                \item If $\delta_i^L = 0$, we have $\delta_i^L - \delta_i' \leq 0$ and $q_i > t_L$. Therefore, $\left\{\frac{1}{1+q_i} - \lambda\left(\frac{q_i}{1+q_i} - \alpha\right)\right\} < \left\{\frac{1}{1+t_L} - \lambda\left(\frac{t_L}{1+t_L} - \alpha\right)\right\} = 0$.
                \item If $\delta_i^L = 1$, we have $\delta_i^L - \delta_i' \geq 0$ and $q_i \leq t_L$. Therefore, $\left\{\frac{1}{1+q_i} - \lambda\left(\frac{q_i}{1+q_i} - \alpha\right)\right\} \geq \left\{\frac{1}{1+t_L} - \lambda\left(\frac{t_L}{1+t_L} - \alpha\right)\right\} = 0$.
            \end{enumerate}
            This proves Equation \eqref{eq:lemma3:4} and concludes the proof.
        \end{proof}

        \subsection{Proof of Lemma \ref{lemma3}}
        
        \begin{proof}
        For $t \geq 0$, we let
        $$
            Q_2(t) = \frac{\sum_{i=1}^{m_2} \{1 - \pi(s_{2,i})\} \hat c_{2,i}(w_{2,i} t)}{ E \{\sum_{i=1}^{m_2} \mathbb{I}(\hat q_{2,i} \leq t)\}},
        $$	
        $$	
        \hat Q_2(t) = \frac{\sum_{i=1}^{m_2} \{1 - \hat \pi_2(s_{2,i})\} \hat c_{2,i}(w_{2,i} t)}{ E \{\sum_{i=1}^{m_2} \mathbb{I}(\hat q_{2,i} \leq t)\}}, 
        $$
        $$
        \widehat{\text{FDP}}_2(t) = \frac{\sum_{i=1}^{m_2} \{1 - \hat \pi_2(s_{2,i})\} \hat c_{2,i}(w_{2,i} t)}{\{\sum_{i=1}^{m_2} \mathbb{I}(\hat q_{2,i} \leq t)\} \lor 1},
        $$
        $$
        t^*_2 = \text{sup} \{ t \in [0, \infty): \widehat{\text{FDP}}(t) \leq \alpha \}.
        $$
        
        As $t^*_2 \geq \hat q_{2,(k)} \geq \nu$ by the first part of Assumption \ref{A9} and Lemma \ref{lemma:eqv}, we only consider $t \geq \nu$ in the following proof. The second part of Assumptiont \ref{A9} implies $ E \{\sum_{i=1}^{m_2}\mathbb{I}(\hat q_{2,i} \leq t)\} \rightarrow \infty$ when $m\rightarrow\infty$ for $t \geq \nu$, which makes $Q_2(t)$, $\hat  Q_2(t)$, $\tilde Q_2(t)$ well defined when $t \geq \nu$.
        
        Note that, by Assumption \ref{A9} and the standard Chernoff bound for independent Bernoulli random variables, we have uniformly for $t \geq \nu$ and any $\epsilon > 0$
        \begin{equation*}
            \begin{split}
                & P  \left( \left| \frac{\sum_{i=1}^{m_2} \mathbb{I}(\hat q_{2,i} \leq t)}{ E \{\sum_{i=1}^{m_2} \mathbb{I}(\hat q_{2,i} \leq t)\}} -1\right| \geq \epsilon \right) \\
                \leq & 2e^{-\epsilon^2\sum_{i=1}^{m}  P (\hat q_{2,i}  \leq t)/3}\\
                \leq&  2e^{-\epsilon^2\sum_{i=1}^{m}  P (\hat q_{2,i}  \leq \nu)/3}
                \rightarrow  0,
            \end{split}
        \end{equation*}
        which implies
        \begin{equation}
        \label{eq:lemma4:1}
            \sup_{t\geq\nu}|\hat Q_2(t) - \widehat{\text{FDP}}_2(t)| \overset{P}{\rightarrow} 0
        \end{equation}
    as $m_2 \rightarrow \infty$.	On the other hand, we have uniformly for $t \geq \nu$,
        \begin{equation*}
            \begin{split}
                |\hat Q_2(t) - \tilde Q_2(t)| 
                =& \left| \frac{\sum_{i=1}^{m_2} \{\tilde \pi_2(s_{2,i}) - \hat \pi_2(s_{2,i})\} \hat c_{2,i}(w_{2,i} t)}{ E \{\sum_{i=1}^{m_2} \mathbb{I}(\hat q_{2,i} \leq t)\}} \right| \\
                \leq& \frac{|\sum_{i=1}^{m_2} \{\tilde \pi_2(s_{2,i}) - \hat \pi_2(s_{2,i})\}|}{ E \{\sum_{i=1}^{m_2} \mathbb{I}(\hat q_{2,i} \leq \nu)\}}  \\
                =& \frac{m_2 \times o_P(1)}{m_2} 
                = o_P(1),
            \end{split}
        \end{equation*}
        where the first $o_P(1)$  is with regard to $m_1 \rightarrow \infty$ by the uniformly conservative consistency of $\hat \pi_2(\cdot)$, and the term $m_2$ in the denominator comes from the first part of Assumption \ref{A9}. As the data splitting strategy ensures $m_1\approx m_2$, we obtain the second $o_P(1)$ with regard to $m \rightarrow \infty$. Thus, we have
        \begin{equation}
        \label{eq:lemma4:2}
            \sup_{t\geq\nu}|\hat Q_2(t) - \tilde Q_2(t)| \rightarrow 0
        \end{equation}
        in probability as $m \rightarrow \infty$. 
        We note that Equation \eqref{eq:lemma4:2} holds in a similar manner for Theorems \ref{thm2} and \ref{dd:wbh}, where their conservative consistency is defined in terms of convergence in probability.
        
        Combining Equations \eqref{eq:lemma4:1} and \eqref{eq:lemma4:2}, we have 
        \begin{equation*}
            \sup_{t\geq\nu}|\widehat{\text{FDP}}_2(t) - \tilde Q_2(t)| \rightarrow 0
        \end{equation*}
        in probability as $m \rightarrow \infty$. Then, following the proof of Theorem \ref{thm4}, we can similarly obtain $t^*_2\rightarrow t_{2,L}$ in probability by Assumptions \ref{A10} and \ref{A11}. Finally, we have
        \begin{equation*}
                \begin{split}
                    \text{mFDR}_{\text{Half-procedure}}
                    & = \frac{\sum_{i=1}^{m_2}  P (\hat q_{2,i} \leq t^*_2, \theta_i = 0)}{ E \{\sum_{i=1}^{m_2} \mathbb{I}(\hat q_{2,i} \leq t^*_2)\}}\\
                    & = \frac{\sum_{i=1}^{m_2}  P (\hat q_{2,i} \leq t_{2,L}, \theta_i = 0)}{ E \{\sum_{i=1}^{m_2} \mathbb{I}(\hat q_{2,i} \leq t_{2,L})\}} + o(1)\\
                    & = Q_2(t_{2,L}) + o(1)
                     \leq \tilde Q_2(t_{2,L}) + o(1)\\
                    & \leq \alpha + o(1).
                \end{split}
            \end{equation*}
        \end{proof}

	\end{appendices}
\newpage
\bibliographystyle{apalike}
\bibliography{reference}

\begin{thebibliography}{}

\bibitem[Aubert et~al., 2004]{aubert2004determination}
Aubert, J., Bar-Hen, A., Daudin, J.-J., and Robin, S. (2004).
\newblock Determination of the differentially expressed genes in microarray experiments using local fdr.
\newblock {\em BMC bioinformatics}, 5(1):1--9.

\bibitem[Basu et~al., 2018]{basu2018weighted}
Basu, P., Cai, T.~T., Das, K., and Sun, W. (2018).
\newblock Weighted false discovery rate control in large-scale multiple testing.
\newblock {\em J. Am. Statist. Assoc.}, 113(523):1172--1183.

\bibitem[Bates et~al., 2023]{bates2023testing}
Bates, S., Cand{\`e}s, E., Lei, L., Romano, Y., and Sesia, M. (2023).
\newblock Testing for outliers with conformal p-values.
\newblock {\em The Annals of Statistics}, 51(1):149--178.

\bibitem[Benjamini and Hochberg, 1995]{benjamini1995controlling}
Benjamini, Y. and Hochberg, Y. (1995).
\newblock Controlling the false discovery rate: a practical and powerful approach to multiple testing.
\newblock {\em J. R. Statist. Soc. B}, 57(1):289--300.

\bibitem[Benjamini and Hochberg, 1997]{benjamini1997multiple}
Benjamini, Y. and Hochberg, Y. (1997).
\newblock Multiple hypotheses testing with weights.
\newblock {\em Scand. J. Stat.}, 24(3):407--418.

\bibitem[Benjamini and Yekutieli, 2001]{benjamini2001control}
Benjamini, Y. and Yekutieli, D. (2001).
\newblock The control of the false discovery rate in multiple testing under dependency.
\newblock {\em Ann. Statist.}, 29(4):1165--1188.

\bibitem[Cai et~al., 2019]{tony2019covariate}
Cai, T.~T., Sun, W., and Wang, W. (2019).
\newblock Covariate-assisted ranking and screening for large-scale two-sample inference.
\newblock {\em J. R. Statist. Soc. B}, 81(2):187--234.

\bibitem[Cai et~al., 2022]{cai2022laws}
Cai, T.~T., Sun, W., and Xia, Y. (2022).
\newblock Laws: A locally adaptive weighting and screening approach to spatial multiple testing.
\newblock {\em J. Am. Statist. Assoc.}, 117(539):1370--1383.

\bibitem[Cao et~al., 2022]{cao2022optimal}
Cao, H., Chen, J., and Zhang, X. (2022).
\newblock Optimal false discovery rate control for large scale multiple testing with auxiliary information.
\newblock {\em Ann. Statist.}, 50(2):807--857.

\bibitem[Chen, 2019]{chen2019uniformly}
Chen, X. (2019).
\newblock Uniformly consistently estimating the proportion of false null hypotheses via lebesgue--stieltjes integral equations.
\newblock {\em J. Multivar. Anal.}, 173:724--744.

\bibitem[Chen and Liu, 2018]{chen2018}
Chen, X. and Liu, W. (2018).
\newblock Testing independence with high-dimensional correlated samples.
\newblock {\em Ann. Statist.}, 46(2):866--894.

\bibitem[Dobriban et~al., 2015]{dobriban2015optimal}
Dobriban, E., Fortney, K., Kim, S.~K., and Owen, A.~B. (2015).
\newblock Optimal multiple testing under a gaussian prior on the effect sizes.
\newblock {\em Biometrika}, 102(4):753--766.

\bibitem[Du and Zhang, 2014]{du2014single}
Du, L. and Zhang, C. (2014).
\newblock Single-index modulated multiple testing.
\newblock {\em Ann. Statist.}, 42(3):1262--1311.

\bibitem[Efron, 2003]{efron2003robbins}
Efron, B. (2003).
\newblock Robbins, empirical bayes and microarrays.
\newblock {\em Ann. Statist.}, 31(2):366--378.

\bibitem[Efron, 2004]{efron2004large}
Efron, B. (2004).
\newblock Large-scale simultaneous hypothesis testing: the choice of a null hypothesis.
\newblock {\em J. Am. Statist. Assoc.}, 99(465):96--104.

\bibitem[Efron and Tibshirani, 2007]{efron2007testing}
Efron, B. and Tibshirani, R. (2007).
\newblock On testing the significance of sets of genes.
\newblock {\em Ann. Appl. Stat.}, 1(1):107--129.

\bibitem[Efron et~al., 2001]{efron2001empirical}
Efron, B., Tibshirani, R., Storey, J.~D., and Tusher, V. (2001).
\newblock Empirical bayes analysis of a microarray experiment.
\newblock {\em J. Am. Statist. Assoc.}, 96(456):1151--1160.

\bibitem[Fu et~al., 2022]{HART}
Fu, L., Gang, B., James, G.~M., and Sun, W. (2022).
\newblock Heteroscedasticity-adjusted ranking and thresholding for large-scale multiple testing.
\newblock {\em J. Am. Statist. Assoc.}, 117(538):1028--1040.

\bibitem[Genovese and Wasserman, 2002]{genovese2002operating}
Genovese, C. and Wasserman, L. (2002).
\newblock Operating characteristics and extensions of the false discovery rate procedure.
\newblock {\em J. R. Statist. Soc. B}, 64(3):499--517.

\bibitem[Genovese et~al., 2006]{genovese2006false}
Genovese, C.~R., Roeder, K., and Wasserman, L. (2006).
\newblock False discovery control with p-value weighting.
\newblock {\em Biometrika}, 93(3):509--524.

\bibitem[Heller and Rosset, 2021]{heller2021optimal}
Heller, R. and Rosset, S. (2021).
\newblock Optimal control of false discovery criteria in the two-group model.
\newblock {\em Journal of the Royal Statistical Society Series B: Statistical Methodology}, 83(1):133--155.

\bibitem[Hong et~al., 2009]{hong2009local}
Hong, W.-J., Tibshirani, R., and Chu, G. (2009).
\newblock Local false discovery rate facilitates comparison of different microarray experiments.
\newblock {\em Nucleic Acids Res.}, 37(22):7483--7497.

\bibitem[Ignatiadis and Huber, 2021]{ignatiadis2021covariate}
Ignatiadis, N. and Huber, W. (2021).
\newblock Covariate powered cross-weighted multiple testing.
\newblock {\em J. R. Statist. Soc. B}, 83(4):720--751.

\bibitem[Jin and Cai, 2007]{jin2007estimating}
Jin, J. and Cai, T.~T. (2007).
\newblock Estimating the null and the proportion of nonnull effects in large-scale multiple comparisons.
\newblock {\em J. Am. Statist. Assoc.}, 102(478):495--506.

\bibitem[Lei and Fithian, 2018]{lei2018adapt}
Lei, L. and Fithian, W. (2018).
\newblock Adapt: an interactive procedure for multiple testing with side information.
\newblock {\em J. R. Statist. Soc. B}, 80(4):649--679.

\bibitem[Leung and Sun, 2022]{ZAP}
Leung, D. and Sun, W. (2022).
\newblock Zap: z-value adaptive procedures for false discovery rate control with side information.
\newblock {\em J. R. Statist. Soc. B}, 84(5):1886--1946.

\bibitem[Li and Barber, 2019]{li2019multiple}
Li, A. and Barber, R.~F. (2019).
\newblock Multiple testing with the structure-adaptive benjamini--hochberg algorithm.
\newblock {\em J. R. Statist. Soc. B}, 81(1):45--74.

\bibitem[Liang et~al., 2023]{liang2023lasla}
Liang, Z., Cai, T.~T., Sun, W., and Xia, Y. (2023).
\newblock Locally adaptive algorithms for multiple testing with network structure, with application to genome-wide association studies.
\newblock {\em arXiv preprint arXiv:2203.11461}.

\bibitem[Liu, 2013]{liu2013gaussian}
Liu, W. (2013).
\newblock Gaussian graphical model estimation with false discovery rate control.
\newblock {\em Ann. Statist.}, 41(6):2948--2978.

\bibitem[Ma et~al., 2023]{ma2022napa}
Ma, L., Xia, Y., and Li, L. (2023).
\newblock {NAPA:} neighborhood-assisted and posterior-adjusted two-sample inference.
\newblock {\em Stat. Sin.}, (just-accepted):1--42.

\bibitem[Marandon et~al., 2024]{marandon2024adaptive}
Marandon, A., Lei, L., Mary, D., and Roquain, E. (2024).
\newblock Adaptive novelty detection with false discovery rate guarantee.
\newblock {\em The Annals of Statistics}, 52(1):157--183.

\bibitem[McDonald et~al., 2018]{mcdonald2018american}
McDonald, D., Hyde, E., Debelius, J.~W., Morton, J.~T., Gonzalez, A., Ackermann, G., Aksenov, A.~A., Behsaz, B., Brennan, C., Chen, Y., et~al. (2018).
\newblock American gut: an open platform for citizen science microbiome research.
\newblock {\em Msystems}, 3(3):10--1128.

\bibitem[Meinshausen and Rice, 2006]{meinshausen2006estimating}
Meinshausen, N. and Rice, J. (2006).
\newblock Estimating the proportion of false null hypotheses among a large number of independently tested hypotheses.
\newblock {\em Ann. Statist.}, 34(1):373--393.

\bibitem[Paloyelis et~al., 2007]{paloyelis2007functional}
Paloyelis, Y., Mehta, M.~A., Kuntsi, J., and Asherson, P. (2007).
\newblock Functional mri in adhd: a systematic literature review.
\newblock {\em Expert review of neurotherapeutics}, 7(10):1337--1356.

\bibitem[Ramdas et~al., 2019]{ramdas2019unified}
Ramdas, A.~K., Barber, R.~F., Wainwright, M.~J., and Jordan, M.~I. (2019).
\newblock A unified treatment of multiple testing with prior knowledge using the p-filter.
\newblock {\em Ann. Statist.}, 47(5):2790--2821.

\bibitem[Roquain and Van De~Wiel, 2009]{roquain2009optimal}
Roquain, E. and Van De~Wiel, M.~A. (2009).
\newblock Optimal weighting for false discovery rate control.
\newblock {\em Electron. J. Stat.}, 3:678--711.

\bibitem[Sarkar and Zhao, 2022]{sarkar2017local}
Sarkar, S.~K. and Zhao, Z. (2022).
\newblock Local false discovery rate based methods for multiple testing of one-way classified hypotheses.
\newblock {\em Electron. J. Stat.}, 16:6043--6085.

\bibitem[Storey, 2002]{storey2002direct}
Storey, J.~D. (2002).
\newblock A direct approach to false discovery rates.
\newblock {\em J. R. Statist. Soc. B}, 64(3):479--498.

\bibitem[Storey et~al., 2007]{storey2007optimal}
Storey, J.~D., Dai, J.~Y., and Leek, J.~T. (2007).
\newblock The optimal discovery procedure for large-scale significance testing, with applications to comparative microarray experiments.
\newblock {\em Biostatistics}, 8(2):414--432.

\bibitem[Sun and Cai, 2007]{sun2007oracle}
Sun, W. and Cai, T.~T. (2007).
\newblock Oracle and adaptive compound decision rules for false discovery rate control.
\newblock {\em J. Am. Statist. Assoc.}, 102(479):901--912.

\bibitem[Vovk and Wang, 2021]{vovk2021values}
Vovk, V. and Wang, R. (2021).
\newblock E-values: Calibration, combination and applications.
\newblock {\em Ann. Statist.}, 49(3):1736--1754.

\bibitem[Wang and Ramdas, 2022]{wang2022false}
Wang, R. and Ramdas, A. (2022).
\newblock False discovery rate control with e-values.
\newblock {\em J. R. Statist. Soc. B}, 84(3):822--852.

\bibitem[Xia, 2020]{xia2020correlation}
Xia, Y. (2020).
\newblock Correlation and association analyses in microbiome study integrating multiomics in health and disease.
\newblock {\em Progress in molecular biology and translational science}, 171:309--491.

\bibitem[Xie et~al., 2011]{xie2011optimal}
Xie, J., Cai, T.~T., Maris, J., and Li, H. (2011).
\newblock Optimal false discovery rate control for dependent data.
\newblock {\em Stat. Interface}, 4(4):417.

\bibitem[Zhang and Chen, 2022]{zhang2022covariate}
Zhang, X. and Chen, J. (2022).
\newblock Covariate adaptive false discovery rate control with applications to omics-wide multiple testing.
\newblock {\em J. Am. Statist. Assoc.}, 117(537):411--427.

\end{thebibliography}

\end{document}